\documentclass[11pt,a4paper,epsf,epsfig,psfrag]{article}
\usepackage{jheppub}
\usepackage{amsmath,graphicx,amsfonts, mathrsfs,amssymb}
\usepackage{amsmath,epsfig}
\usepackage{amssymb,amsfonts}
\usepackage{color}
\usepackage{latexsym}
\usepackage{epsfig}
\usepackage{pdfsync}
\usepackage{marvosym}
\usepackage{mathtools}
\usepackage[utf8]{inputenc}
\newbox\pippobox

\def \be {\begin{equation}}
	\def \ee {\end{equation}}
\def \ba {\begin{array}}
	\def \ea {\end{array}}
\def \bea{\begin{eqnarray}}
	\def \eea{\end{eqnarray}}
\def \nn {\nonumber}

\def \m {\mu}
\def \n {\nu}

\def \r {\rho}

\def \p {\partial}
\def \f {\frac}

\def \sr {\sqrt}

\def \inf {\infty}

\usepackage{cancel,soul,ulem}
\title{\textbf{Signals of critical end point from jet quenching and quark energy loss in holographic QCD}}

\author[a,1]{Zhibin Li,}
\author[b,2]{Danning Li,}
\author[c,3]{and Mei Huang}

\affiliation[a]{Institute for Astrophysics, School of Physics, Zhengzhou University, Zhengzhou 450001, China}
\affiliation[b]{Department of Physics and Siyuan Laboratory, Jinan University, Guangzhou 510632, P.R. China}
\affiliation[c]{School of Nuclear Science and Technology, University of Chinese Academy of Sciences, Beijing 100049, China}
\emailAdd{lizhibin@zzu.edu.cn}
\emailAdd{lidanning@jnu.edu.cn}
\emailAdd{huangmei@ucas.ac.cn}

\date{\today}

\abstract{The jet quenching parameter and energy loss of light and heavy quarks have been investigated in the framework of holographic QCD models with a critical end point (CEP) at finite baryon chemical potential in $N_f=2,2+1,2+1+1$ systems. The properties of the jet quenching parameter and energy loss around CEP have been carefully studied, and some evident signatures for CEP are subtracted. It is found that the dimensionless jet quenching parameter and the energy loss of light and heavy quarks exhibit evident features around CEP. Specifically, all these quantities increase rapidly near the CEP phase transition temperature $T_{CEP}$ with fixed $\mu_{CEP}$. Moreover, the velocity dependent behavior of heavy quark energy loss at CEP differs significantly from charged particle energy loss in  QED matter. For electromagnetic interaction, the energy loss of charged particle can be described by the Bethe-Bloch formula and the Lindhand-Scharff-Schiott theory at low and high velocities, respectively. However, the heavy quark energy loss at CEP is approximately proportional to velocity at low velocities and aligns with Bjorken's results at high velocities, which indicates that the heavy quark energy loss is predominantly collisional at low velocities and  gluon radiation dominant at high velocities. For light quark energy loss, the behavior of the energy loss per unit length and the total energy loss  differs significantly. However, the total energy loss and stopping distance exhibit similar behavior. This implies that the stopping distance predominantly determines the total energy loss. Thus, even with increased energy loss per unit length at higher temperatures or chemical potentials, the total energy loss decreases due to the reduced stopping distance. }

\begin{document}

\maketitle

%%%%%%%%%%%%%%%%%%%%%%%%%%%%%%%%%%%%%%%%%%%%%%
\section{Introduction}\label{sec:01}
%%%%%%%%%%%%%%%%%%%%%%%%%%%%%%%%%%%%%%%%%%%%%%
It is expected that a new state of matter, the deconfined quark-gluon plasma (QGP), can be created through heavy-ion collisions, and it has been an important target of the beam energy scan program of heavy-ion collisions to explore the Quantum Chromodynamics (QCD) phase diagram at finite temperature and baryon chemical potential, especially to locate the critical end point (CEP). The latest STAR results \cite{Ashish2024,STAR:2025zdq} on net baryon number fluctuations $\kappa \sigma^2$ did not provide clear evidence for the location of CEP with collision energy $\sqrt{s}>$ 7.7 GeV, which highlights the need to explore other signatures of CEP that are not affected by freeze-out \cite{Shen:2024ptz}. Jet quenching and energy loss might be a good candidate.

An effective approach to study the properties of QGP is to investigate the energy loss experienced by high-momentum particles originating from high-energy partons as they interact with the hot and dense medium. The jet quenching parameter $\hat{q}$ quantifies the average transverse momentum squared transferred per unit length from the medium to a fast-moving parton as \cite{Baier:1996sk}
\begin{equation*}
    \hat{q} = \left\langle \frac{d p_\perp^2}{d L} \right\rangle,
\end{equation*}
where $p_\perp$ is the transverse momentum acquired by the parton, and $L$ is the path length traversed in the medium. It governs jet quenching, a phenomenon in which jets lose energy and broaden in transverse momentum due to interactions with the medium. Observables such as jet suppression ($R_{\text{AA}}$) and azimuthal anisotropy in heavy-ion collisions are directly linked to $\hat{q}$ \cite{Shan-Liang:2023sdk}. 

The JET Collaboration has analyzed the experimental data of nuclear modification factors for hadron spectra at RHIC and LHC, based on various jet quenching models, and has extracted the dependence of QGP jet quenching parameters on the initial temperature \cite{JET:2013cls}.  Near the phase boundary, the jet quenching parameter has been observed to exhibit a peak-like structure \cite{Li:2014hja,Xu:2014tda,Du:2022oaw} or to undergo a rapid increase with temperature \cite{JET:2013cls}. Lattice QCD studies have investigated the values of jet-quenching parameters at zero density. For example, simulations within the framework of electrostatic QCD (EQCD) have shown that the soft contribution to the jet quenching parameter $\hat{q}$ is significantly larger than predicted by perturbative methods at experimentally relevant temperatures. These studies highlight the importance of non-perturbative effects in understanding jet quenching phenomena in heavy-ion collisions \cite{Panero:2013pla}. Additionally, Bayesian parameter estimation methods have been employed to constrain the jet quenching parameter \cite{JETSCAPE:2021ehl,Xie:2022ght}. The jet quenching parameter can also be used to probe CEP, for example, through the phenomenon of partial subcritical opalescence \cite{Wu:2022vbu}.

The holographic calculation of the jet quenching parameters is based on the expectation values of light-like rectangular Wilson loops described by quark-antiquark pairs \cite{Liu:2006ug,Liu:2006he}. This approach leverages the AdS/CFT correspondence, where the Wilson loop in a strongly coupled plasma can be mapped to a string configuration in the dual gravitational background. The expectation value of the Wilson loop provides insight into the interaction potential between quark-antiquark pairs and, consequently, the energy loss experienced by high-energy partons propagating through the medium \cite{Li:2014hja,Li:2014dsa,Zhu:2020qyw,Cao:2024jgt,Chen:2022goa}. The behavior of jet quenching parameters has been extensively studied under extreme conditions that may occur in heavy-ion collision experiments, such as the presence of strong magnetic fields \cite{Rougemont:2020had} and rotational conditions \cite{Chen:2022obe}. These studies show that the jet quenching parameter increases with magnetic field strength and rotational velocity, indicating enhanced energy loss in these environments.

In relativistic heavy-ion collisions, high-energy partons generated in the initial hard scattering will lose energy through various processes such as collisions and radiation \cite{Ficnar:2014pmc} when passing through the heat and dense QGP material produced after the collision. The law governing the energy loss of high-energy partons moving in hot QCD media is of interest to both theoretical and experimental researchers. It is observed that as the momentum of heavy partons increases, the energy loss increases dramatically from smaller values and then tends to saturate \cite{Jacobs:2004qv,YousufJamal:2019pen}. Currently, there is no universal law that governs the energy loss of heavy quarks. This is mainly due to the fact that heavy-quark energy loss involves multiple mechanisms and is influenced by a variety of factors. These factors include the properties of the medium through which the quarks move—such as temperature and density—as well as the kinematic properties of the heavy quarks themselves.

The drag force, which quantifies the resistance experienced by a heavy quark moving through the plasma, is directly related to energy loss and can be calculated from the rate of energy loss per unit distance. In holographic research, people use the tension on a classical string with one end connected to the boundary of a 5-dimensional bulk spacetime to describe the drag force experienced by heavy partons moving in the medium at the 4-dimensional boundary spacetime \cite{Gubser:2006bz,Herzog:2006gh,Gursoy:2009kk,Rougemont:2015wca,Zhu:2021nbl,Grefa:2022sav,Zhu:2023aaq,Chen:2024epd,Domurcukgul:2021qfe,Finazzo:2016mhm}. 

To study the energy loss caused by the motion of light quarks in a medium, it is essential to consider quarks with finite momentum as probes. The energy loss of the light quark can be calculated using the falling string \cite{Gubser:2008as,Chesler:2008wd,Ficnar:2012yu} or the shooting string method \cite{Ficnar:2013qxa,Ficnar:2013wba}. In this work, we will employ the shooting string method to compute the energy loss of light quarks. This method involves modeling the dynamics of a classical string in a holographic background, where the endpoint of the string initially carries energy and momentum. These quantities are gradually transferred to the rest of the string as it evolves \cite{Ficnar:2013qxa,Ficnar:2013wba}. In the presence of a chemical potential or a magnetic field, the energy loss of light quarks is enhanced, which is consistent with the findings from drag force calculations and jet quenching parameters \cite{Rougemont:2015wca,Zhu:2019ujc}. 

With the integration of machine learning and other innovative approaches, holographic QCD models have achieved remarkable progress in quantification \cite{Chen:2024ckb,Chen:2024epd,Chen:2024mmd,Cai:2024eqa,Fu:2024wkn}. Over the past several years, significant efforts have been devoted to applying gauge/gravity duality to elucidate the properties of QCD matter. Within the soft-wall model \cite{Karch:2006pv}, researchers have successfully described the properties of QCD matter at both zero and finite temperatures by employing a quadratic dilaton field. This includes studies on chiral and deconfinement phase transitions \cite{Li:2011hp,Li:2012ay,Chelabi:2015gpc,Chelabi:2015cwn,Fang:2015ytf,Fang:2019lsz}. The Einstein-Maxwell-Dilaton (EMD) system, which incorporates a bulk non-conformal dilatonic scalar and a $U(1)$ gauge field, has been widely utilized in holographic QCD studies~\cite{ DeWolfe:2010he, DeWolfe:2011ts, Cai:2012xh, Cai:2012eh, Finazzo:2013efa, Yang:2014bqa, Critelli:2017oub, Li:2017ple, Chen:2017cyc, Knaute:2017opk, Fang:2018axm, Ballon-Bayona:2020xls, Li:2020hau, Grefa:2021qvt, He:2022amv, Grefa:2022fpu, Chen:2024ckb}. In this work, we employ an EMD model \cite{Chen:2024mmd} constructed by machine learning, featuring analytical forms of metrics and thermodynamic quantities, to investigate the jet quenching parameter and the energy loss of partons. Transport coefficients and jet energy loss of hot and dense quark matter have been studied in holographic models
\cite{Grefa:2022sav,Grefa:2023hmf} with a non-conformal dilaton potential. In this work, we investigate the features of jet quenching and energy loss around CEP in the Einstein-Maxwell dilaton holographic framework with different flavors \cite{Chen:2024ckb,Chen:2024epd,Chen:2024mmd}. 

The rest of this article is organized as follows. In Section 2, we provide a brief introduction to the holographic model and the parameter settings used. In Section 3, we present the general derivation and numerical results for the calculation of the jet quenching parameter within the holographic framework. In Section 4, we derive and present numerical results for the heavy-quark energy loss. In Section 5, we derive and present numerical results for the light-quark energy loss. Finally, Section 6 concludes with a summary and discussion of our findings.

%%%%%%%%%%%%%%%%%%%%%%%%%%%%%%%%%%%%%%%%%%%%%%
\section{The Einstein-Maxwell-dilaton holographic QCD model}\label{sec:02}
%%%%%%%%%%%%%%%%%%%%%%%%%%%%%%%%%%%%%%%%%%%%%%
We focus on a 5-dimensional Einstein-Maxwell-dilaton holographic model with the action \cite{He:2013qq,Yang:2014bqa,DeWolfe:2010he,Li:2017tdz}
\be
S=\f{1}{16\pi G_5} \int d^5 x \sr{-g}\left[R-\f{f\left(\phi\right)}{4}F_{\m\n}^2-\f{1}{2}(\p\phi)^2-V(\phi)\right],
\ee
with $G_5$ the five-dimensional Newton's constant, $F_{\mu\nu}$ the field strength of a $U(1)$ gauge field $A_\mu$ dual to the conserved current of baryon number, $\phi$ the dilaton field and $V(\phi)$ of its potential, and $f(\phi)$ the gauge kinetic function. 
The ansatz of the metric is taken as
\be\label{eq22}
ds^2=\f{e^{2A(z)}}{z^2}[-g(z)dt^2+\f{1}{g(z)}dz^2+d\vec{x}^2],
\ee
with the regular boundary conditions at the horizon $z=z_h$ and the asymptotic $AdS_5$ condition at the boundary $z=0$ \cite{Yang:2014bqa,Li:2017tdz}
\be
A_t(z_h)=g(z_h)=0,
\ee
\be
A(0)=-\sr{\f{1}{6}}\phi(0), \hspace{0.3cm} g(0)=1,
\ee
and the UV expansion of gauge field $A_t(z)$ as
\be
A_t(z\rightarrow 0)=\m+8\pi G_5\r z^2+\cdots,
\ee
where $\m$ and $\r$ are the chemical potential and density of the quark, respectively. The gauge kinetic function can be fixed as 
\be
f(\phi(z))=e^{c z^2-A(z)+k},
\ee
in which two free parameters $c,k$ are introduced and the function $A(z)$ will be fixed later.
Then the analytical solution of equation of motion takes the following form:
\begin{equation}
\begin{aligned}
g(z)&=1-\frac{1}{\int_0^{z_h} d x x^3 e^{-3 A(x)}}\left[\int_0^z d x x^3 e^{-3 A(x)}+\frac{2 c \mu^2 e^k}{\left(1-e^{-c z_h^2}\right)^2} \operatorname{det} \Gamma \right],\\
\phi^{\prime}(z) & =\sqrt{6\left(A^{\prime 2}-A^{\prime \prime}-2 A^{\prime} / z\right)}, \\
A_t(z) & =\mu \frac{e^{-c z^2}-e^{-c z_h^2}}{1-e^{-c z_h^2}}, \\
V(z) & =-\frac{3 z^2 g e^{-2 A}}{L^2}\left[A^{\prime \prime}+A^{\prime}\left(3 A^{\prime}-\frac{6}{z}+\frac{3 g^{\prime}}{2 g}\right)-\frac{1}{z}\left(-\frac{4}{z}+\frac{3 g^{\prime}}{2 g}\right)+\frac{g^{\prime \prime}}{6 g}\right],
\end{aligned}
\end{equation}
in which the matrix $\Gamma$ takes the following form
\begin{equation}
\Gamma =\left[\begin{array}{ll}
\int_0^{z_h} d y y^3 e^{-3 A(y)} & \int_0^{z_h} d y y^3 e^{-3 A(y)-c y^2} \\
\int_{z_h}^z d y y^3 e^{-3 A(y)} & \int_{z_h}^z d y y^3 e^{-3 A(y)-c y^2}
\end{array}\right].
\end{equation}
Then from the equation of motion we can calculate the quark density $\r$ and temperature $T$ as
\be
\r=\rho(z_h,\mu)=\frac{c \mu e^{A(0)} f(\phi (0))}{8\pi G_5 \left(1-e^{-c z_h^2}\right)},
\ee
\be
T=T(z_h,\mu)=\frac{z_h^3 e^{-3 A\left(z_h\right)}}{4 \pi \int_0^{z_h} d y y^3 e^{-3 A(y)}}\left[1+\frac{2 c \mu^2 e^k\left(e^{-c z_h^2} \int_0^{z_h} d y y^3 e^{-3 A(y)}-\int_0^{z_h} d y y^3 e^{-3 A(y)} e^{-c y^2}\right)}{\left(1-e^{-c z_h^2 }\right)^2}\right].
\ee
The entropy density $s$ can be calculated as \cite{DeWolfe:2011ts}
\be
s=s(z_h)=\f{e^{3A(z_h)}}{4 G_5 z_h^3},
\ee
and the free energy density can be calculated by integral
\bea
F&=&F(z_h,\mu)=-\int \left(s d T+\rho d \mu\right)\\ \nn
&=&\int_{z_h}^\infty s(\tilde{z}_h)\frac{\partial T(\tilde{z}_h,\mu)}{\partial \tilde{z}_h} d \tilde{z}_h-\int_0^\mu \left(\frac{\partial T(z_h,\tilde{\mu})}{\partial \tilde{\mu}}  +\rho(z_h,\tilde{\mu})\right) d \tilde{\mu}+F(\inf,0).\\
\eea
Following \cite{Chen:2024ckb,Chen:2024epd,Chen:2024mmd}, the warped factor $A(z)$ can be fixed as 
\be
A(z)=d \ln (a z^2+1)+d\ln (b z^4+1),
\ee
with three additional parameters $a,b,d$ introduced.

In these settings, in the previous study \cite{Chen:2024mmd}, it predicts continuous crossovers at low baryon number densities and first-order phase transitions at high densities, while critical end points with temperatures and chemical potentials $T=T_{CEP}, \mu=\mu_{CEP}$ are shown to appear between them. Here we adopt the parameters given in Table \ref{table:parameter}, which are fitted through machine learning in \cite{Chen:2024mmd}. Under all these settings, one can easily obtain the background metric (semi-)analytically or numerically. In later calculations, we will apply those solutions directly. For more details of these solutions, please refer to Ref. \cite{Chen:2024mmd}.
\begin{table}[htbp]
	\centering
	\begin{tabular}{|c|c|c|c|c|c|c|c|}
		\hline
		$N_f$  & $a$ & $b$ & $c$ & $d$ & $k$ & $G_5$ &  CEP (GeV) \\
        \hline
		 2 & 0.067 & 0.023 & -0.377 & -0.382 & 0 & 0.885 &  ($\mu_{CEP}$=0.46, $T_{CEP}$=0.147) \\
        \hline
        2+1 & 0.204 & 0.013 & -0.264 & -0.173 & -0.824 & 0.400 & ($\mu_{CEP}$=0.74, $T_{CEP}$=0.094) \\
        \hline
       2+1+1 & 0.196 & 0.014 & -0.362 & -0.171 & -0.735 & 0.391 & ($\mu_{CEP}$=0.87, $T_{CEP}$=0.108) \\
        \hline
	\end{tabular}
\caption{Parameters given by the machine learning of $2$ flavor, $2+1$ flavor system, and $2+1+1$ flavor system, respectively \cite{Chen:2024mmd}. $\mu_{CEP}$ and $T_{CEP}$ are the baryon chemical potential and temperature of CEP, respectively.}
\label{table:parameter}
\end{table}

\section{Jet Quenching Parameter}
Holographic calculation of the jet quenching parameter $\hat{q}$ is related to specially configured Wilson loops as
\begin{equation}
W^{Adj}_{C} \approx \exp\left(-\frac{1}{4\sqrt{2}} \hat{q} L^{-} L^2\right), \label{eq30}
\end{equation}
where $W^{Adj}_C$ is the Wilson loop in the adjoint representation. $C$ is given by the trajectories of quarks and antiquarks moving along light-like curves on the boundary. The quark and antiquark are separated in the $x_2$ direction by a small distance $L$ and travel $L^-$ along the $x^-=(t-x_1)/\sqrt{2}$ direction.

Assuming that the spacetime metric in string frame has the following form
\begin{equation}
    ds^2=-g_{tt} dt^2+g_{zz}dz^2+g_{xx}d\vec{x}^2
\end{equation}
with $d\vec{x}^2=dx_1^2+dx_2^2+dx_3^2$ and $g_{tt}$, $g_{zz}$ and $g_{xx}$ all functions of $z$, the metric in the light-cone coordinates $x^{\pm}=(t\pm x_1)/\sqrt{2}$ has the following form
\begin{equation}
    ds^2=-\frac{1}{2}\left(g_{tt}-g_{xx}\right) \left[d(x^+)^2+d(x^-)^2\right]-\left(g_{tt}+g_{xx}\right) dx^+dx^-+g_{zz}dz^2+g_{xx}\left(dx_2^2+dx_3^2\right).
\end{equation}
The action on the string world sheet is as follows
\begin{equation}
    S_{NG}=\frac{1}{2 \pi \alpha'}\int d\sigma d\tau \sqrt{-\det \gamma_{\alpha_\beta}},
\end{equation}
with $\gamma_{\alpha\beta}=g_{\mu\nu} \partial_\alpha x^\mu \partial_\beta x^\nu$ 
the induced metric on the string world sheet. And $\alpha,~\beta=1,~2$  with 
$\partial_1=\partial/\partial \tau$ and $\partial_2=\partial/\partial \sigma$ as well as $\tau=x^-$, $\sigma=x_2$. At the limit $L^-\gg L$ the classic configurations of string is translational invariant along the $\tau$ direction. The action can be calculated as
\begin{equation}
    S_{NG}=\frac{L^-}{2 \pi \alpha'}\int d\sigma \sqrt{-\det \gamma_{\alpha\beta}}.
\end{equation}
The renormalized action can be given as $ S_R=S_c-S_d$, with $S_c$ and $S_d$ the action of connected and disconnected configurations of string, respectively. Then the adjoint Wilson loop can be calculated as
\begin{equation}
W^{Adj}_{C}=e^{-2 S_R}.
\end{equation}
Thus, one has
\begin{equation}
\hat{q}=\frac{8\sqrt{2}}{L^-L^2}S_R,
\end{equation}
since $\hat{q}$ is proportional to the coefficient of $L^2$ in $S_R$ at the small $L$ limit. The connected configuration of string can be defined as $z=z(x_2)$ and the disconnected one with $x_2=\pm L/2$. Thus the actions take the forms of
\begin{eqnarray} 
   S_c&=&\frac{L^-}{2 \pi \alpha'}\int_{-L/2}^{L/2} dx_2 \sqrt{\frac{1}{2}\left(g_{tt}-g_{xx}\right)\left(g_{xx}+g_{zz}z'(x_2)^2 \right)}  \\ \nn
   &=&\frac{L^-}{ \pi \alpha'}\int_{0}^{L/2} dx_2 \sqrt{\frac{1}{2}\left(g_{tt}-g_{xx}\right)\left(g_{xx}+g_{zz}z'(x_2)^2 \right)},  \\ \nn
   S_d&=&\frac{L^-}{ \pi \alpha'}\int_0^{z_h} dz \sqrt{\frac{1}  {2}g_{zz}\left(g_{tt}-g_{xx}\right)}.
\end{eqnarray}
The constant of motion for the connected action $S_c$ can be easily derived as 
\begin{equation}
    \mathcal{H}=z'(x_2) \frac{\partial \mathcal{L}}{\partial z'(x_2)}-\mathcal{L}=g_{xx}\sqrt{\frac{g_{tt}-g_{xx}}{2(g_{xx}+z'(x_2)^2 g_{zz})}},
\end{equation}
where $\mathcal{L}=\sqrt{\frac{1}{2}\left(g_{tt}-g_{xx}\right)\left(g_{xx}+g_{zz}z'(x_2)^2 \right)}=\frac{1}{2\mathcal{H}}g_{xx}\left(g_{tt}-g_{xx}\right)$. Correspondingly, one can get
\begin{equation}
    \frac{dx_2}{dz}=z'(x_2)^{-1}=\pm \sqrt{2}\mathcal{H} \sqrt{\frac{g_{zz}}{g_{xx}\left[ g_{xx}(g_{tt}-g_{xx})-\mathcal{H}^2\right]}}.
\end{equation}
The zero point of $dx_2/dz$ gives the position of the midpoint of the string in the $z$ direction which is at $z=z_h$. And the distance between quark and anti-quark is
\begin{equation}
    L=2\int_0^{z_h} \frac{1}{z'(x_2)}dz=2\sqrt{2}\mathcal{H}\int_0^{z_h}dz \sqrt{\frac{g_{zz}}{g_{xx}\left[ g_{xx}(g_{tt}-g_{xx})-\mathcal{H}^2\right]}}.
\end{equation}
To obtain the small $L$ expansion of $S_R$, it is convenient to expand all quantities on the small $\mathcal{H}$. When $L\ll L^-$, we have
\begin{equation}\label{311}
L=2\sqrt{2}\mathcal{H}\int_0^{z_h}dz \frac{1}{g_{xx}}\sqrt{\frac{g_{zz}}{g_{tt}-g_{xx}}}+\mathcal{O}(\mathcal{H}^3)
\end{equation}
and 
\begin{equation}
    S_c=\frac{L^-}{ \pi \alpha'}\int_{0}^{L/2} dx_2\mathcal{L}=\frac{L^-}{ \pi \alpha'}\int_{0}^{z_h} dz\frac{1}{z'(x_2)}\mathcal{L}=\frac{L^-}{ \pi \alpha'}\int_{0}^{z_h} dz\sqrt{\frac{g_{xx}g_{zz}(g_{tt}-g_{xx})^2}{2 g_{xx} (g_{tt}-g_{xx})-4\mathcal{H}^2}}.
\end{equation}
And the small $L$ expansion takes the form
\begin{eqnarray}
     S_c&=&\frac{L^-}{ \pi \alpha'}\int_0^{z_h} dz \sqrt{\frac{1}  {2}g_{zz}\left(g_{tt}-g_{xx}\right)}+\frac{L^-\mathcal{H}^2}{2 \pi \alpha'}\int_0^{z_h} dz\frac{1}{g_{xx}} \sqrt{\frac{2 g_{zz}}  {g_{tt}-g_{xx}}}+\mathcal{O}(\mathcal{H}^3).
\end{eqnarray}
Therefore, for small $L$ we have 
\begin{equation}
    S_R=S_c-S_d=\frac{L^-\mathcal{H}^2}{2 \pi \alpha'}\int_0^{z_h} dz\frac{1}{g_{xx}} \sqrt{\frac{2 g_{zz}}  {g_{tt}-g_{xx}}}+\mathcal{O}(\mathcal{H}^3).
\end{equation}
By using Eq. obtain
\begin{equation}
    S_R=\frac{L^-L^2}{8 \pi \alpha'}\left[\int_0^{z_h} dz\frac{1}{g_{xx}} \sqrt{\frac{2 g_{zz}}  {g_{tt}-g_{xx}}}\right]^{-1}+\mathcal{O}(L^3).
\end{equation}
Finally it is easy to get a simple form
\begin{equation}
\hat{q}=\frac{8\sqrt{2}}{L^-L^2}S_R=\frac{\sqrt{2}}{\pi \alpha'}\frac{1}{\int_0^{z_h} dz\frac{1}{g_{xx}} \sqrt{\frac{2 g_{zz}}  {g_{tt}-g_{xx}}}}.
\end{equation}
\begin{figure}[htbp]
\centering
\includegraphics[width=.48\textwidth]{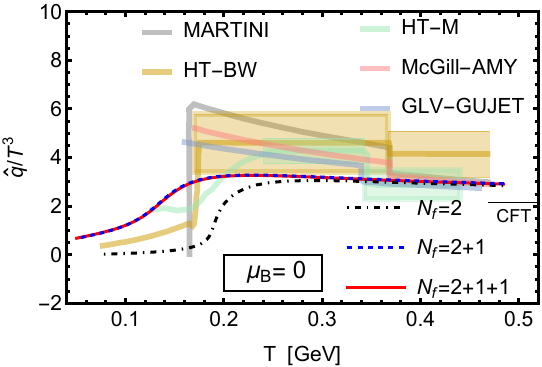}
\includegraphics[width=.48\textwidth]{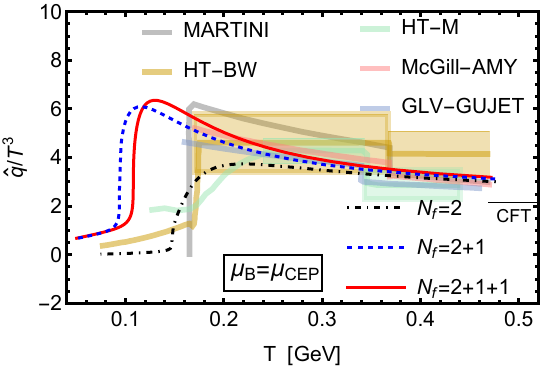}
\caption{Temperature dependence of jet quenching parameter at $\mu_B=0$ (\textbf{Left panel}) and $\mu_B=\mu_{CEP}$ (\textbf{Right panel}) for $N_f=2,~2+1,~2+1+1$ compared with the results of JET Collaboration in different jet quenching models for a quark jet with initial energy $E = 10~\text{GeV}$ \cite{JET:2013cls}. }
\label{fig1}
\end{figure}

In order to make a quantitative comparison with the data from JET Collaboration \cite{JET:2013cls}, we take the metric solution described in Sec.\ref{sec:02} and set $\alpha'=3.5$ in the different systems: $N_f=2,~2+1,~2+1+1$. And $\sqrt{\lambda_t}=R_c^2/\alpha'$ \cite{Liu:2006ug} with $R_c$ the radius of curvature of the bulk spacetime, which we set to be one as in Eq. \eqref{eq22}. Then we obtain the temperature dependence of the scaled jet quenching parameter $\hat{q}/T^3$ at both vanishing ($\mu_B=0$) and critical ($\mu_B=\mu_{CEP}$) baryon chemical potentials, as shown in Fig.~\ref{fig1}. Comparison of our calculations for different flavor number scenarios with the JET Collaboration model-dependent predictions for initial quark jets of energy $E = 10~\text{GeV}$ \cite{JET:2013cls} are shown as well. The numerical results reveal a characteristic nonmonotonic temperature dependence: $\hat{q}/T^3$ exhibits a pronounced enhancement followed by gradual suppression near the crossover transition region, manifesting as a peak structure. Although all flavor configurations demonstrate qualitatively similar scaling trends with temperature, quantitative disparities emerge primarily in the transition temperature.

\begin{figure}[htbp]
\centering
\includegraphics[width=.4\textwidth]{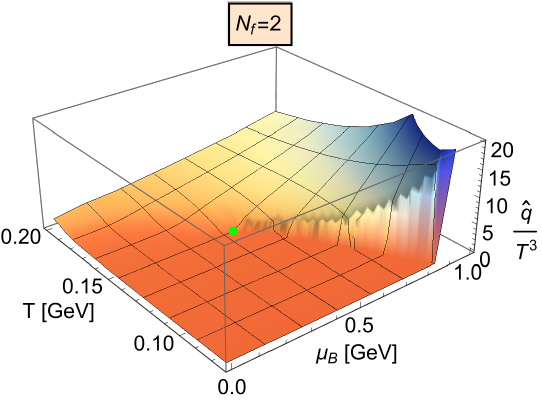}
\qquad
\includegraphics[width=.4\textwidth]{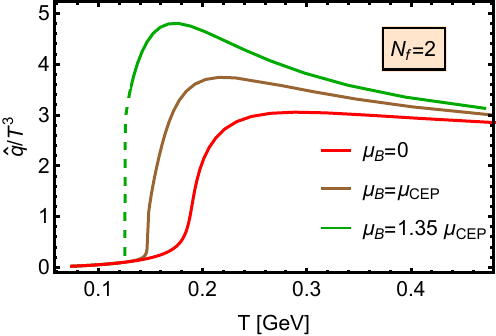}\\
\includegraphics[width=.4\textwidth]{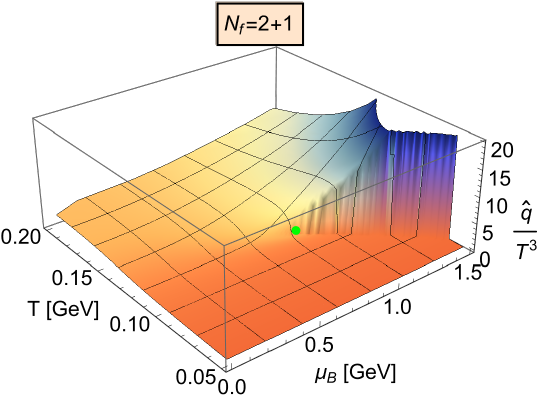}
\qquad
\includegraphics[width=.4\textwidth]{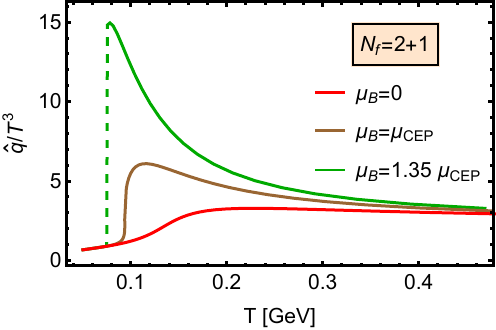}\\
\includegraphics[width=.4\textwidth]{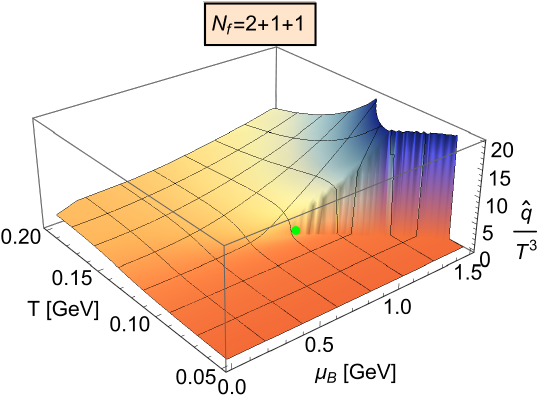}
\qquad
\includegraphics[width=.4\textwidth]{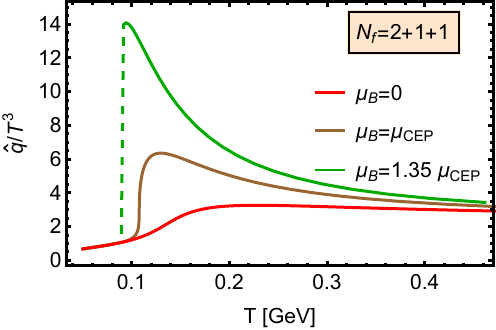}
\caption{Jet quenching parameter $\hat{q}/T^3$ at finite temperature and baryon chemical potential. The results, from top to bottom, correspond to $N_f=2, 2+1, 2+1+1$ flavors, respectively. \textbf{Left panel}: the behavior of jet quenching parameter $\hat{q}/T^3$ in the phase diagram. The CEP is denoted by green dots. \textbf{Right panel}: the temperature dependence of jet quenching parameters at fixed chemical potentials. The red, brown, and green lines correspond to $\mu_B=0,~ \mu_{CEP}~ \text{and}~ 1.35\mu_{CEP}$, respectively. The dashed lines in the figure indicate discontinuous jumps.}\label{fig2}
\end{figure}
%%%

In Fig.~\ref{fig2}, we present the behavior of the normalized jet quenching parameter $\hat{q}/T^3$ in the phase diagram. The 3D plot reveals that $\hat{q}/T^3$ remains at very low values in regions of low temperature and low density. As both the temperature and the chemical potential increase, $\hat{q}/T^3$ exhibits a rapid rise near the phase boundary. At a fixed temperature, $\hat{q}/T^3$ increases monotonically with increasing chemical potential, indicating the enhancement of the interaction between the partons and the medium by high densities. If fixing  $\mu_B$, $\hat{q}/T^3$ initially increases with temperature, then decreases, forming a peak near the phase boundary. Moreover, the peak increases with increasing $\mu_B$. In the high-temperature region, $\hat{q}/T^3$ gradually approaches the conformal limit \cite{Liu:2006he,Liu:2006ug}
\begin{equation}
    \hat{q}_{CFT}=\frac{\pi^{3/2}\Gamma(\frac{3}{4})}{\Gamma(\frac{5}{4})}\sqrt{\lambda_t} T^3
\end{equation}
In the crossover region, $\hat{q}/T^3$ changes smoothly with temperature and baryon chemical potential. At the CEPs, which are denoted by green dots in the left panels, $\hat{q}/T^3$  remains continuous but exhibits   sharp, nearly vertical increase behavior. In contrast, at the first-order phase transition, $\hat{q}/T^3$ undergoes a significant discontinuous change.

\section{Energy loss of heavy quark}
To consider the energy loss of a heavy quark, we consider a quark probe on the boundary moving at a constant velocity $v$. The drag force of a trailing string can describe the energy loss of the quark when it moves through a hot and dense medium \cite{Gubser:2006bz,Herzog:2006gh,Gursoy:2009kk,Rougemont:2015wca,Grefa:2022sav}. The configuration of the trailing string is as follows: one end moving along with the quark and the other end located on the black hole horizon. The trailing string is described by the Nambu-Goto action
\begin{equation}
    S_{NG}=\frac{1}{2 \pi \alpha'}\int d\sigma d\tau \sqrt{-\det \gamma_{\alpha\beta}}
\end{equation}
with $\gamma_{\alpha\beta}=g_{\mu\nu} \partial_\alpha x^\mu \partial_\beta x^\nu$ 
the induced metric on the string world sheet. And $\alpha,~\beta=0,~1$  with 
$\partial_0=\partial/\partial \tau$ and $\partial_1=\partial/\partial \sigma$ as well as $\tau=t$, $\sigma=z$, slightly different from the previous case since the worldsheet embedding is now $x=x(t,z)$. Then the action is
\begin{equation}
    S_{NG}=\int dt dz \mathcal{L}=\frac{1}{2 \pi \alpha'}\int dt dz \sqrt{g_{zz}\left[g_{tt}-g_{xx}(\partial_0 x(t,z))^2\right]+g_{tt}g_{xx}(\partial_1 x(t,z))^2}.
\end{equation}
The worldsheet current of spacetime energy-momentum carried by the string is 
\begin{equation}
    P_\mu^\alpha=-\frac{1}{2\pi \alpha'} \sqrt{-\det \gamma} \gamma^{\alpha\beta}g_{\mu\nu}\partial_\beta x^\nu
\end{equation}
and the equation of motion is
\begin{equation}
    \partial_\alpha P_\mu^\alpha-\Gamma^\kappa_{\mu\lambda}\partial_\beta x^\lambda P^\beta_\kappa=0.
\end{equation}
Under the background described in Sec.\ref{sec:02}, the equation of motion becomes
\begin{equation}
    \partial_0 P^0_0+\partial_1 P^1_0=0, ~~\partial_0 P^0_2+\partial_1 P^1_2=0.
\end{equation}
Then the energy and momentum in the $x$ direction in the bulk of the string can be derived as 
\begin{equation}
    P_x=\int_0^{z_h}dz P_2^0, ~~ E=-\int_0^{z_h}dz P_0^0.
\end{equation}
Then we have
\begin{equation}
    \frac{d P_x}{dt}=\frac{d}{dt}\int_0^{z_h}dz P_2^0=\int_0^{z_h}dz \frac{d}{dt}P_2^0=-\int_0^{z_h}dz \frac{d}{dz}P_2^1=P_2^1|_{z_h}^0
\end{equation}
and 
\begin{equation}
    \frac{d E}{dx}=-\frac{d}{dx}\int_0^{z_h}dz P_0^0=-\int_0^{z_h}dz \frac{dt}{dx} \frac{d}{dt}P_0^0-\frac{dz}{dx}\frac{d}{dz}\int_0^{z_h}dz P_0^0=\int_0^{z_h}dz \frac{dt}{dx}\frac{d}{dz}P_0^1.
\end{equation}
Here we have used the relation
\begin{equation}
    \frac{d P_x}{dt}=\frac{d}{dt}\int_0^{z_h}dz P_2^0=\int_0^{z_h}dz \frac{d}{dt}P_2^0=-\int_0^{z_h}dz \frac{d}{dz}P_2^1=\left.P_2^1\right|_{z_h}^0.
\end{equation}
We choose the stationary trailing string ansatz $x(t,z)=v t +\xi(z)$. Then the Lagrangian becomes a function of $z$ only, i.e.,
\begin{equation}
    \mathcal{L}=\frac{1}{2\pi \alpha' }\sqrt{g_{zz}\left[g_{tt}-g_{xx}v^2\right]+g_{tt}g_{xx}\xi'^2},
\end{equation}
where $\xi'=d\xi(z)/dz$. The drag force acting on the probe quark on the boundary should be
\begin{equation}
    F_{drag}=\frac{d P_{x0}}{dt}=-\left.\frac{dP_x}{dt}\right|_{z=0}=-P_2^1(z=0)
\end{equation}
and the energy loss of the probe quark takes the form of
\begin{equation}
    \frac{d E_{0}}{dX}=-\left.\frac{dE}{dx}\right|_{z=0}=\frac{1}{v}P_0^1(z=0).
\end{equation}
Then we get 
\begin{equation}
    F_{drag}=\frac{d P_{x0}}{dt}= \frac{d E_{0}}{dX}=\left. \frac{g_{tt}g_{xx}\xi'}{2\pi \alpha' \sqrt{g_{zz}\left[g_{tt}-g_{xx}v^2\right]+g_{tt}g_{xx}\xi'^2}}\right|_{z=0}.
\end{equation}
From the equation of motion of string action the conjugate momentum of $\xi$ is a conserved quantity which can be read as
\begin{equation}
    \Pi^x=\frac{\partial \mathcal{L}}{\partial \xi'}=\frac{g_{tt}g_{xx}\xi'}{2\pi \alpha' \sqrt{g_{zz}\left[g_{tt}-g_{xx}v^2\right]+g_{tt}g_{xx}\xi'^2}}
\end{equation}
with $\partial \Pi^x/\partial z=0$. So the drag force becomes
\begin{equation}
    F_{drag}=\left. \frac{g_{tt}g_{xx}\xi'}{2\pi \alpha' \sqrt{g_{zz}\left[g_{tt}-g_{xx}v^2\right]+g_{tt}g_{xx}\xi'^2}}\right|_{z=0}=\frac{g_{tt}g_{xx}\xi'}{2\pi \alpha' \sqrt{g_{zz}\left[g_{tt}-g_{xx}v^2\right]+g_{tt}g_{xx}\xi'^2}}.
\end{equation}
The $\xi'$ can be solved as
\begin{equation}
    \xi'=-\sqrt{\frac{g_{zz}(g_{tt}-v^2g_{xx})}{g_{tt}g_{xx}\left[-1+\frac{1}{4\pi^2\alpha'^2 {\Pi^x}^2}g_{tt}g_{xx}\right]}}.
\end{equation}
Here the negative root corresponds to the string trailing behind the endpoint. Then for a fixed velocity $v$ the relation $dx/dz=\xi'$ requires that $\xi'$ is real in the range of $z \in [0,z_h]$. It means for some $z^\star$ the numerator $g_{zz}(z^\star)(g_{tt}(z^\star)-v^2g_{xx}(z^\star))=0$ the denominator should also equal to $0$, i.e.,
\begin{equation}
    g_{tt}(z^\star)g_{xx}(z^\star)\left[-1+\frac{1}{4\pi^2\alpha'^2 {\Pi^x}^2}g_{tt}(z^\star)g_{xx}(z^\star)\right]=0.
\end{equation}
Then we have
\begin{equation}
  {\Pi^x}^2= \frac{1}{4\pi^2\alpha'^2} g_{tt}(z^\star)g_{xx}(z^\star)
\end{equation}
and
\begin{equation}
  F_{drag}=\frac{d P_{x0}}{dt}= \frac{d E_{0}}{dX}=\Pi^x=-\frac{1}{2\pi\alpha'}\sqrt{ g_{tt}(z^\star)g_{xx}(z^\star)}.
\end{equation}

Up to now, it is not difficult to get the corresponding energy loss of heavy quarks in the holographic models described in Sec.\ref{sec:02}. Here, we will show the numerical results. We will directly normalize the drag force using temperature square $T^2$, without utilizing its conformal limit
\begin{equation}\label{fdragcft}
    F_{drag}^{CFT}=-\frac{\pi}{2}\sqrt{\lambda_t} \frac{v}{\sqrt{1-v^2}}T^2.
\end{equation}

\begin{figure}[htbp]
\centering
\includegraphics[width=.32\textwidth]{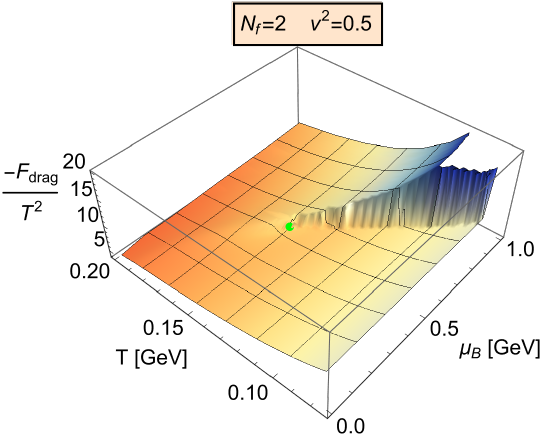}
\includegraphics[width=.32\textwidth]{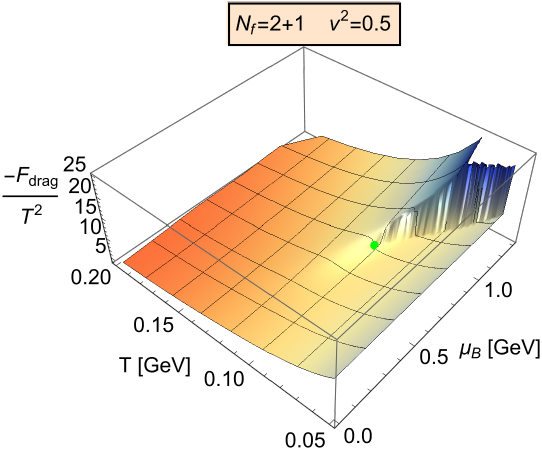}
\includegraphics[width=.32\textwidth]{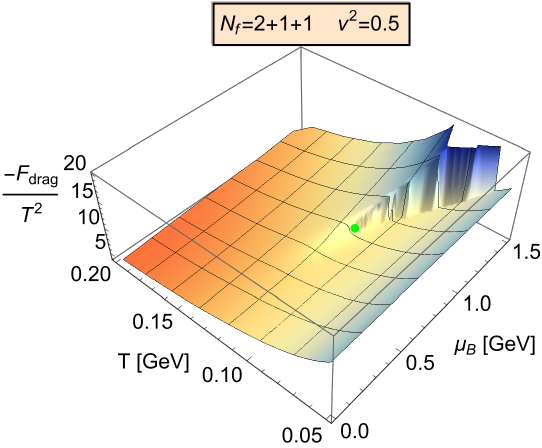}\\
\includegraphics[width=.32\textwidth]{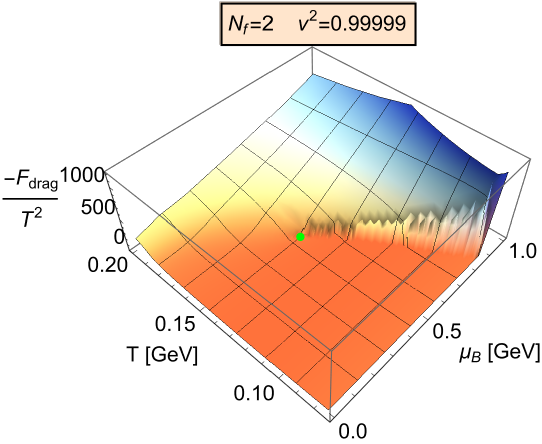}
\includegraphics[width=.32\textwidth]{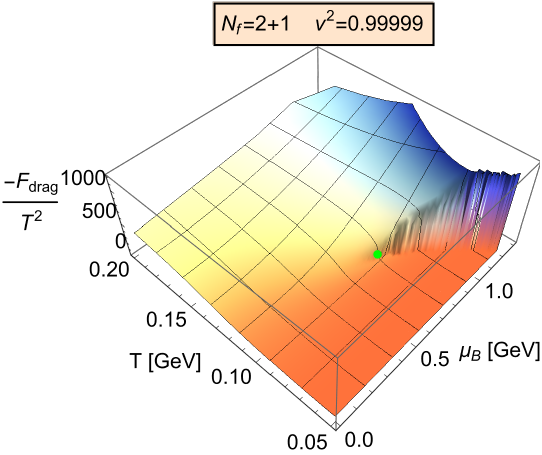}
\includegraphics[width=.32\textwidth]{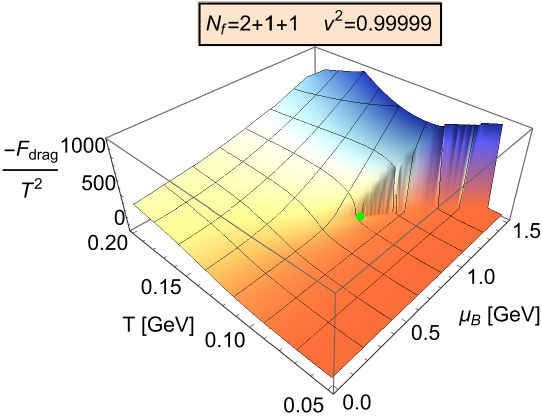}
\caption{Energy loss of heavy quark $F_{drag}/T^2$ at finite temperature and baryon chemical potential for different flavors. The images sequentially correspond to flavor configurations of $N_f=2, 2+1, 2+1+1$ from left to right. The upper and lower panels represent velocities of $v^2=0.5~\text{and}~0.99999$, respectively. The CEP is denoted by green dots.}\label{fig5}
\end{figure}
%%%

Fig.~\ref{fig5} shows the results of the models for $N_f=2,~2+1,~2+1+1$, respectively. Our findings reveal that the drag force, for various flavor models, exhibits a similar behavior across the phase diagram when evaluated at the same velocity. Furthermore, we find that within a broad range of velocities ($v^2\le 0.9$), the drag force shows consistent characteristics. Note that, to facilitate a clear representation of the magnitude of the drag force, the term "drag force" used in the subsequent discussion refers specifically to its absolute value.

\begin{figure}[htbp]
\centering
\includegraphics[width=.32\textwidth]{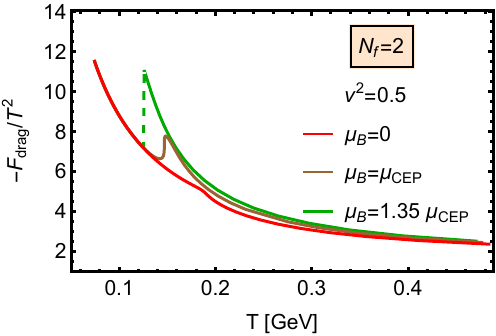}
\includegraphics[width=.32\textwidth]{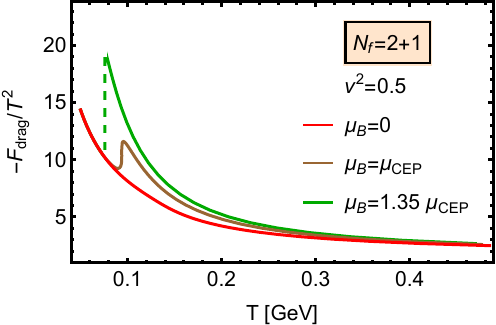}
\includegraphics[width=.32\textwidth]{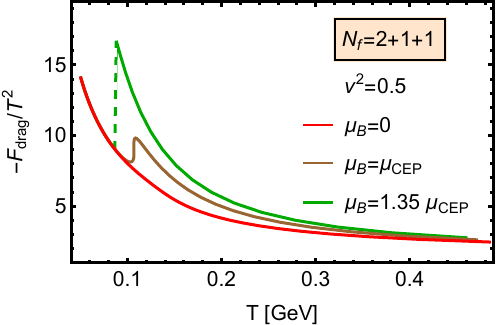}\\
\includegraphics[width=.32\textwidth]{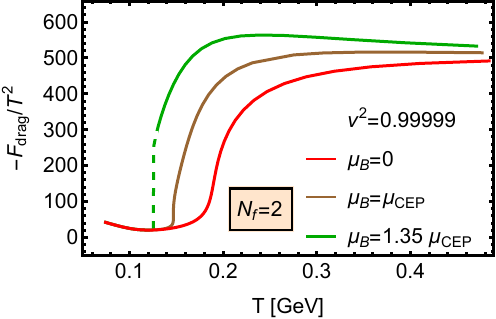}
\includegraphics[width=.32\textwidth]{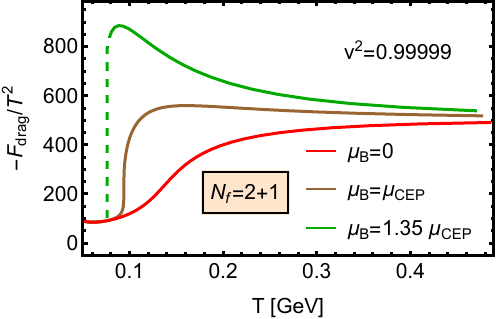}
\includegraphics[width=.32\textwidth]{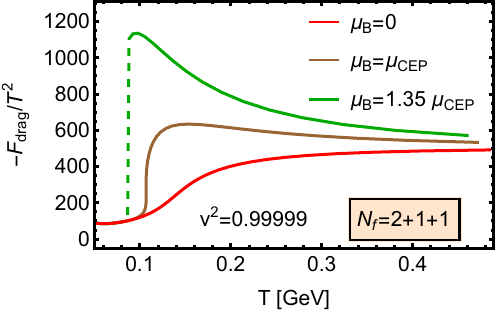}
\caption{The temperature dependence of heavy quark energy loss $F_{drag}/T^2$ at fixed baryon chemical potential for different flavors. The images sequentially correspond to flavor configurations of $N_f=2, 
 2+1, 2+1+1$ from left to right.  The red, brown, and green lines correspond to $\mu_B=0,~ \mu_{CEP}~ \text{and}~ 1.35\mu_{CEP}$, respectively. The dashed lines in the figure indicate discontinuous jumps. The upper and lower panels represent velocities of $v^2=0.5~\text{and}~0.99999$, respectively.}\label{fig6}
\end{figure}

For $v^2\le 0.9$, the drag force is found to be relatively insensitive to changes in chemical potential at both low and high temperatures, while it decreases monotonically with temperature. However, in the vicinity of the phase boundary, the drag force manifests a complex nonmonotonic behavior that is also clearly evident in the 2d plot presented in Fig.~\ref{fig6}. Specifically, in regions with lower chemical potentials, the drag force decreases with temperature. As the chemical potential increases, the behavior of the drag force near the phase boundary transitions from decreasing with temperature to increasing, and then reverts to a decreasing trend after crossing the phase boundary. Near the critical endpoint (CEP), the drag force exhibits a pronounced increase. In the region of the first-order phase transition, the drag force exhibits a characteristic behavior as a function of temperature. Specifically, with increasing temperature, the drag force initially declines, undergoes a discontinuous drop to a significantly higher value at the phase boundary, and then resumes a decreasing trend at higher temperatures. Ultimately, it asymptotically approaches the conformal limit at sufficiently high temperatures. In other words, when $\mu_B$ approaches or exceeds $\mu_{CEP}$ the energy loss is significantly enhanced near the phase boundary, which is consistent with the mean field results in \cite{Wu:2022vbu}. 

In the extreme relativistic regime ($v^2 = 0.99999$), as depicted in the lower panel of Fig.~\ref{fig5} and Fig.~\ref{fig6}, the behavior of the drag force deviates significantly from that observed in the scenario $v^2 \leq 0.9$. For low chemical potentials, the drag force increases with increasing temperature, exhibiting a minor increase only at very low temperatures. As the system approaches the critical end point, the drag force in the high-temperature phase remains almost unchanged with temperature. At high $\mu_B$, the drag force in the high-temperature phase shows a pronounced decrease with temperature. At high temperatures, the drag forces for different chemical potentials converge toward the conformal limit. Additionally, the drag force behavior under various chemical potentials is found to be similar at low temperature. Near the phase boundary, the drag force undergoes a rapid increase. In the crossover region, this increase is smooth, while at the CEP, it manifests itself as a nearly vertical rise. At the first-order phase transition, a significant discontinuous rise is observed. Similar situations have also been found in \cite{Zhu:2023aaq}.

Additionally, for a fixed temperature, the drag force exhibits a monotonic increase with rising chemical potential. The conformal limit, which varies with velocity, also demonstrates an increasing trend as velocity increases, which is consistent with the dependence of the conformal limit on speed Eq. \eqref{fdragcft}.

%%%

%
\begin{figure}[htbp]
\centering
\includegraphics[width=.32\textwidth]{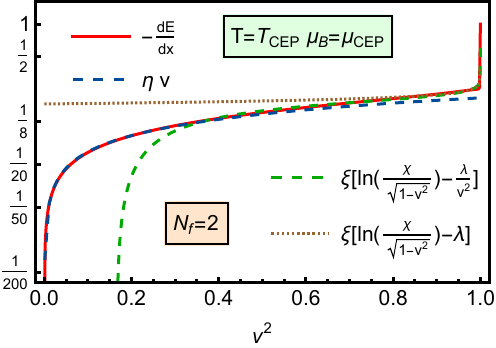}
\hspace{-0.2cm}
\includegraphics[width=.32\textwidth]{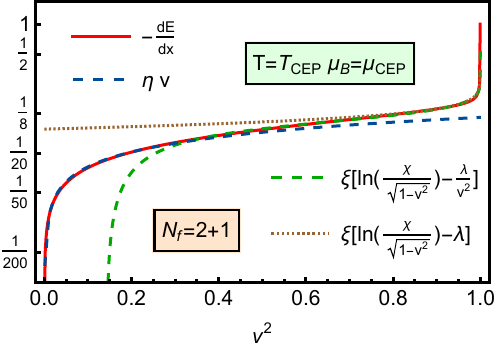}
\hspace{-0.2cm}
\includegraphics[width=.32\textwidth]{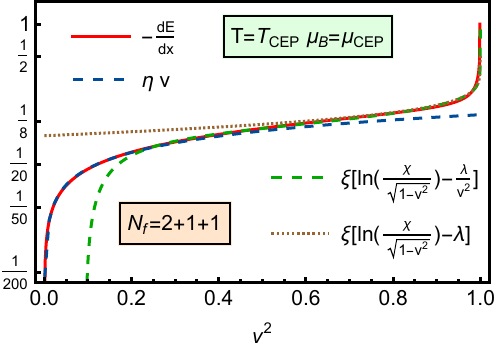}
\caption{Velocity dependent behavior of heavy quark $-dE/dx$ at CEP. The figures, from left to right, correspond to $N_f=2,~2+1,~\text{and} ~2+1+1$, respectively. The red solid line shows the result from holographic QCD calculations. The green dashed line represents the fit obtained using Eq.~\eqref{eq421}. The blue dashed line corresponds to the fit from Eq.~\eqref{eq422}. The brown dotted line indicates the behavior at high velocities described by Eq.~\eqref{eq423}. The parameters in Eq.~\eqref{eq421} for models with different flavors are presented in Tab.~\ref{table2}.}\label{fig8}
\end{figure}
%%%
\begin{table}[htbp]
	\centering
	\begin{tabular}{|c|c|c|c|c|}
		\hline
		$N_f$  & $\xi$ $[\text{GeV}^2]$ & $\chi$ & $\lambda$ & $\eta$ $[\text{GeV}^2]$\\
        \hline
		 2 & 0.035 & 374.303 & 0.950 &  0.208 \\
        \hline
        2+1 & 0.046 & 9.305 & 0.330 &  0.115  \\
        \hline
       2+1+1 & 0.074 & 4.030 & 0.136 &  0.144  \\
        \hline
	\end{tabular}
\caption{Parameters in Eq.~\eqref{eq421} by fitting the energy loss data from holographic QCD models for $2$ flavor, $2+1$ flavor, and $2+1+1$ flavor, respectively. }
\label{table2}
\end{table}
Finally, in Fig.~\ref{fig8} we demonstrate the relationship between drag force $F_{drag}=dE/dx$ and incident particle velocity at CEP in different flavor models. The relationship between energy loss and velocity exhibited by different flavor models is highly similar, characterized by a monotonic increase with increasing velocity. More precisely, the energy loss of heavy quarks in a medium is found to exhibit an approximate velocity dependence
\begin{equation}\label{eq421}
   \left. -\frac{dE}{dx}\right|_{v^2\geq 0.3}=\xi\left[ \ln\left( \frac{\chi}{\sqrt{1-v^2}} \right) -\frac{\lambda}{v^2}\right]=\xi \left[\ln \left(\chi \gamma\right)-\frac{\lambda}{\beta^2} \right]=\xi \left[\ln \left(\frac{\chi E}{M}\right)-\frac{\lambda}{\beta^2} \right]
\end{equation}
at high velocity $v^2\geq 0.3$ which is shown as the green dashed line in the Fig.\ref{fig8}.  Here, $\xi$, $\chi$ and $\lambda$ are parameters that depend on the temperature and chemical potential. We use the symbols $\beta=v/c$, $\gamma=1/\sqrt{1-v^2/c^2}$, and $c=1$ and the relation $E=\gamma M$. As for the low-velocity region, we obtain a different behavior
\begin{equation}\label{eq422}
    \left. -\frac{dE}{dx}\right|_{v^2\leq 0.3}=\eta v,
\end{equation}
as shown by blue dashed line in the Fig.\ref{fig8}. As indicated by the green dashed line in Fig.\ref{fig8}, in the relativistic limit, the behavior of drag force has the form of
\begin{equation}\label{eq423}
   \left. -\frac{dE}{dx}\right|_{v\approx c}=\xi\left[ \ln\left( \frac{\chi}{\sqrt{1-v^2}} \right) -\lambda\right]=\xi \left[\ln \left(\frac{\chi E}{\text{e}^\lambda M}\right)\right],
\end{equation}
which is consistent with Bjorken's results \cite{Bjorken:1982tu}
\begin{equation}
     -\frac{dE}{dx}=2\pi \alpha_s^2 \left( 1+\frac{N_f}{6} \right) \left(\frac{2}{3}\right)^{\pm 1} T^2 \ln \left(\frac{4 T E}{M^2}\right),
\end{equation}
with $\alpha_s$, $N_f$ and $T$ the coupling constant, number of flavor and temperature, respectively. The $+$ ($-$) is for quarks (gluons).  As shown in Tab.~\ref{table2}, the parameters for different quark flavors in the CEP can be numerically extracted. It shows that the coefficient $\xi$ increases with increasing $N_f$, qualitatively aligning with Bjorken's formula. 

Interestingly, the energy loss of partons moving in a QGP medium shows a velocity dependence similar to that of charged particles in an electromagnetic medium. For charged particles, the energy loss at high velocities follows the Bethe-Bloch formula (BBF) \cite{SCHMIDT1987287} that can be parameterized as
\begin{equation}
     -\frac{dE}{dx}=\frac{1}{c_1 \beta^2}\left[\ln \left( c_2 \beta \gamma\right) -\beta^2-\frac{c_3}{2} \right]
\end{equation}
with $c_1,~c_2,~\text{and}~c_3$ parameters related to the properties of charged particles and electromagnetic medium. At low velocities, as predicted by the Lindhand-Scharff-Schiott theory (LSST), the energy loss $dE/dx \propto v$ \cite{Lindhard:1961zz}. 

\begin{figure}[htbp]
\centering
\includegraphics[width=.48\textwidth]{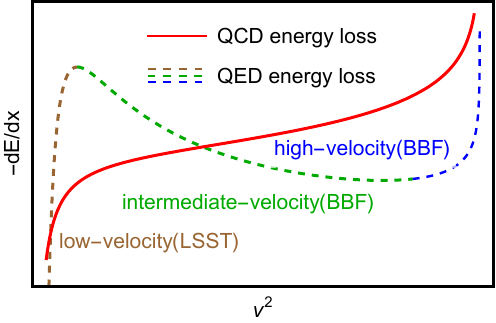}
\caption{Sketch map of the comparison of the characteristic behavior of energy loss for charged particles in electromagnetic media, derived from the Lindhand-Scharff-Schiott theory (LSST) \cite{Lindhard:1961zz} and Bethe-Bloch formula (BBF) \cite{SCHMIDT1987287}, and the energy loss of particles in strongly interacting, thermally dense media obtained from holographic QCD. The dashed lines depict energy loss due to electromagnetic interactions, with different colors corresponding to various velocity ranges. The solid red line indicates energy loss due to strong interactions.}\label{figbbf}
\end{figure}

The energy loss of charged particles in electromagnetic media, linked to mechanisms such as collision, ionization, and radiation, is well understood. In contrast, the energy-loss mechanisms of partons in QGP media remain unclear. The similarity in the velocity - dependent energy loss behaviors of these two scenarios allows for analysis to identify the dominant energy loss mechanisms and influencing factors for partons interacting with QGP at different velocities. Based on the Bethe-Bloch formula and Lindhand-Scharff-Schiott theory, the characteristic behavior of energy loss for charged particles in electromagnetic media as a function of the square velocity is shown in Fig.~\ref{figbbf}. At low velocities, charged particles often capture electrons from the medium, becoming electrically neutral. This neutralization reduces their electromagnetic interaction with the medium. Consequently, energy loss is mainly due to collisions with atomic nuclei, a phenomenon known as nuclear stopping \cite{Bohr:1941zz}. Within this range, the energy loss is proportional to the velocity that can be characterized by the Lindhand-Scharff-Schiott theory (LSST) \cite{Lindhard:1961zz}. Similar collision processes can also occur during interactions between partons and the medium \cite{Stoecker:1994ud,NA49:1998gaz}. In the high-velocity regime, energy loss is dominated by relativistic effects, such as bremsstrahlung and Cherenkov radiation in electromagnetic medium and gluon radiation in QCD medium. These processes follow similar trends, because of the relativistic nature of the interactions, which enhances the energy loss rate with increasing velocity.

However, as shown in Fig.~\ref{figbbf} within the intermediate velocity range, the behavior of energy loss differs significantly between the electromagnetic and QCD mediums. For charged particles in an electromagnetic medium, the energy loss of charged particles within this velocity range is primarily due to electron ionization and excitation. In this velocity regime, the energy loss of charged particles, primarily from electromagnetic interactions with extranuclear electrons, is proportional to the interaction time and thus proportional to $1/v$. In contrast, we did not observe the phenomenon of energy loss decreasing with increasing velocity when studying the motion of partons in the QCD medium. This difference may arise because strong interactions lack the ionization and excitation effect present in electromagnetic interactions.

In summary, at low velocities, the energy loss is primarily due to collisions between partons and medium particles, whereas at high velocities, the energy loss is predominantly attributed to the radiation of gluons.

%%%

\section{Energy loss of the light quark}
To calculate the energy loss of light quark, the light quark should move in the bulk but not be constrained on the boundary. We need consider an open string with the ends take finite momentum which can be described by the action \cite{Ficnar:2013wba,Ficnar:2013qxa}
\begin{equation}
    S_{NG}=\frac{1}{2 \pi \alpha'}\int_M d\sigma^0 d\sigma^1 \sqrt{-\det h}h^{ab}g_{\mu\nu}\partial_a x^\mu\partial_b x^\nu+\int_{\partial M}d\xi \frac{1}{2\eta}g_{\mu\nu}\dot{x}^\mu\dot{x}^\nu,
\end{equation}
where $\xi$ is an arbitrary way to parametrize the boundary of string and dots represent the differentiation with respect to $\xi$. Here $x^\mu$ is the space-time coordinate and $\partial_a$ is the derivative of the worldsheet coordinate $\sigma^a$. $h_{ab}$ is the worldsheet metric with
\begin{equation}
    h_{ab}=\partial_a x^\mu \partial_b x^\nu g_{\mu\nu}.
\end{equation}
The field $\eta$ is an auxiliary field, and its equation of motion is
\begin{equation}
    \dot{x}^\mu \dot{x}^\nu g_{\mu\nu}=0,
\end{equation}
which gives string endpoints moving at the speed of light in the bulk. The current of the string in the bulk is
\begin{equation}
    P_\mu^\alpha=-\frac{1}{2\pi \alpha'} \sqrt{-\det h} h^{\alpha\beta}g_{\mu\nu}\partial_\beta x^\nu
\end{equation}
and the energy momentum on the boundary is
\begin{equation}\label{eq55}
    p_\mu=\frac{1}{\eta}g_{\mu\nu} \dot {x}^\nu.
\end{equation}
The equation of motion in the bulk and on the boundary gives
\begin{equation}
    \partial_a P_\mu^a-\Gamma^\kappa_{\mu\lambda}\partial_b x^\lambda P^b_\kappa=0,~~~~~~ \dot{p}_\mu-\Gamma^\kappa_{\mu\lambda}\dot{x}^\lambda p_\kappa=\mp\frac{\eta}{2\pi \alpha'}p_\mu=\mp \frac{1}{2\pi \alpha'} g_{\mu\nu}\dot{x}^\nu.
\end{equation}
In the last "$=$" we have used the relation in Eq. \eqref{eq55}. Here, $\pm$ are for the shooting and falling strings, respectively. We will take the shooting string which moves from the horizon with the boundary conditions $z=z_h$ and $x=0$. Then we have
\begin{equation}
\dot{p}_\mu=\Gamma^\kappa_{\mu\lambda}\dot{x}^\lambda p_\kappa-\frac{1}{2\pi \alpha'} g_{\mu\nu}\dot{x}^\nu=\Gamma^\kappa_{\mu\lambda}\dot{x}^\lambda \frac{1}{\eta}g_{\kappa\nu} \dot {x}^\nu-\frac{1}{2\pi \alpha'} g_{\mu\nu}\dot{x}^\nu
\end{equation}
and
\begin{equation}
\dot{p}_0=-\frac{1}{2\pi \alpha'} g_{0\nu}\dot{x}^\nu=\frac{1}{2\pi \alpha'} g_{tt}\dot{t}, ~~~~~~\dot{p}_2=-\frac{1}{2\pi \alpha'} g_{2\nu}\dot{x}^\nu=-\frac{1}{2\pi \alpha'} g_{xx}\dot{x},
\end{equation}
which gives
\begin{equation}
\dot{E}=\frac{1}{2\pi \alpha'} g_{tt}\dot{t}, ~~~~~~\dot{P}_x=-\frac{1}{2\pi \alpha'} g_{xx}\dot{x}.
\end{equation}
It is not difficult to prove that the ratio of endpoint energy to momentum is a constant
\begin{equation}
R=\frac{E}{P_x}=\frac{p_0}{p_2}=\frac{g_{0\nu} \dot{x}^\nu}{g_{2\rho} \dot{x}^\rho}=-\frac{g_{tt} \dot{t}}{g_{xx} \dot{x}}
\end{equation}
with
\begin{equation}
\dot{R}=\frac{d}{d\xi}\frac{p_0}{p_2}=\frac{\dot{p}_0p_2-p_0\dot{p}_2}{p_2^2}=\frac{1}{p_2^2}\left[\frac{1}{2\pi \alpha'} g_{tt}\dot{t}\frac{1}{\eta}g_{xx}\dot{x}-\left(-\frac{1}{2\pi \alpha'} g_{xx}\dot{x}\right)\left(-\frac{1}{\eta}g_{tt}\dot{t}\right)\right]=0.
\end{equation}
The light-like equation $\dot{x}^\mu \dot{x}^\nu g_{\mu\nu}=0$ then becomes
\begin{equation}
     -g_{tt}^{-1}\left(R g_{xx}\dot{x}\right)^2+g_{zz}\dot{z}^2+g_{xx}\dot{x}^2=0.
\end{equation}
Using the relation $\dot{x}/\dot{z}=dx/dz$, we have 
\begin{equation}\label{geoeq}
    \left(\frac{dx}{dz}\right)^2=\frac{g_{zz}g_{tt}}{g_{xx}\left(R^2 g_{xx}-g_{tt}\right)},
\end{equation}
which is the orbital equation for the motion of the string endpoint in the $x-z$ plane. The value of $R$ is related to the bounce point $z_\star$ of the string with 
\begin{equation}
\left. \dot{z}\right|_{z=z_\star}=-\left.\sqrt{\frac{g_{xx}\left(R^2 g_{xx}-g_{tt}\right)}{g_{zz}g_{tt}}}\dot{x}\right|_{z=z_\star}=0
\end{equation}
and 
\begin{equation}
R=\sqrt{\frac{g_{tt}(z_\star)}{g_{xx}(z_\star)}}.
\end{equation}
At last we have the energy loss
\begin{equation}
\frac{dE}{dx}=\frac{1}{2\pi \alpha'} g_{tt}\frac{dt}{dx}=-\frac{1}{2\pi \alpha'} g_{tt}\frac{R g_{xx}}{g_{tt}}=-\frac{1}{2\pi \alpha'}R g_{xx}, ~~~~~~\frac{d P_x}{dx}=-\frac{1}{2\pi \alpha'} g_{xx},
\end{equation}
and the velocity
\begin{equation}
\frac{dx}{dt}=-\frac{g_{tt}}{R g_{xx}}.
\end{equation}

In this section, we show the results of the energy loss of the light quark. We normalized the energy loss of light quark by its conformal limit \cite{Ficnar:2013qxa}
\begin{equation}
    \frac{dE_{CFT}}{dx}=-\frac{\pi}{2}\sqrt{\lambda_t}\left( 1+\pi T x \right)^2 T^2
\end{equation}
and we also set $\alpha'=\sqrt{\lambda_t}=1$ in this section. Here we define the total energy loss of the shooting string
\begin{equation}
    \Delta E=\int dE=\int \frac{dE}{dx}dx=\int_{z_h}^{z_\star}\frac{dE}{dx}\frac{dx}{dz}dz=-\frac{1}{2\pi\alpha'}R \int_{z_h}^{z_\star} g_{xx}\sqrt{\frac{g_{zz} g_{tt}}{g_{xx}\left( R^2 g_{xx}-g_{tt}\right)}}dz.
\end{equation}
And the stopping distance at the bounces point can be defined as
\begin{equation}
    \Delta x=\int dx=\int_{z_h}^{z_\star}\frac{dx}{dz}dz=\int_{z_h}^{z_\star} \sqrt{\frac{g_{zz} g_{tt}}{g_{xx}\left( R^2 g_{xx}-g_{tt}\right)}}dz .
\end{equation}
Note that $\Delta E$ represents the total energy loss of the string from the moment of emission until it bounces back. With the solution of Eq. \eqref{geoeq} for AdS-Schwarzschild spacetime, it is also easy to get the conformal limit as
\begin{equation}
    \Delta E_{CFT}=\int_0^{x(z_\star)} \frac{dE_{CFT}}{dx}dx=\frac{1}{6}\sqrt{\lambda_t}\left[1-\left( 1+\pi T x(z_\star) \right)^3\right] T
\end{equation}
and 
\begin{equation}
   x(z)=\frac{1}{\pi T} \left[\frac{1}{\pi Tz}\prescript{}{2} F _1\left( \frac{1}{4},\frac{1}{2},\frac{5}{4},\frac{z_\star^4}{z^4} \right)- \prescript{}{2}  F_1\left( \frac{1}{4},\frac{1}{2},\frac{5}{4},\pi^4 T^4 z_\star^4 \right) \right],
\end{equation}
where $\prescript{}{2}  F_1$ is the ordinary hypergeometric function.

\begin{figure}[htbp]
\centering
\includegraphics[width=.4\textwidth]{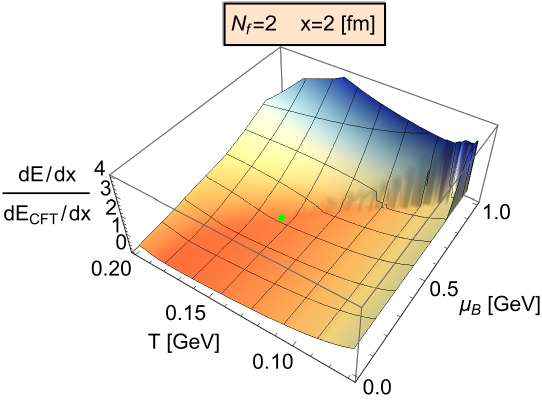}
\hspace{0.1cm}
\includegraphics[width=.42\textwidth]{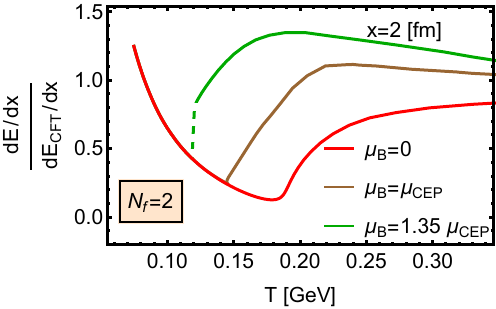}\\
\includegraphics[width=.4\textwidth]{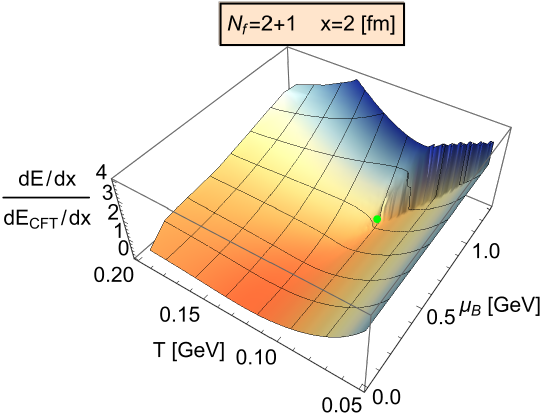}
\hspace{0.3cm}
\includegraphics[width=.4\textwidth]{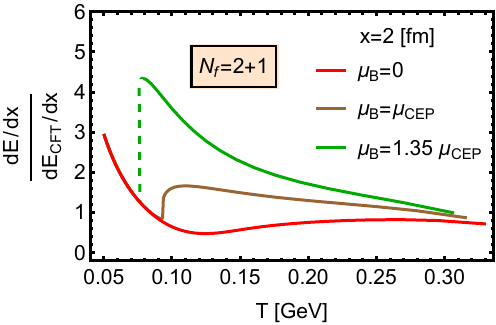}\\
\includegraphics[width=.4\textwidth]{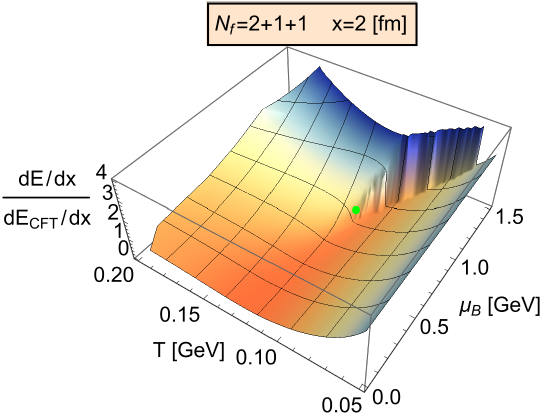}
\qquad
\includegraphics[width=.4\textwidth]{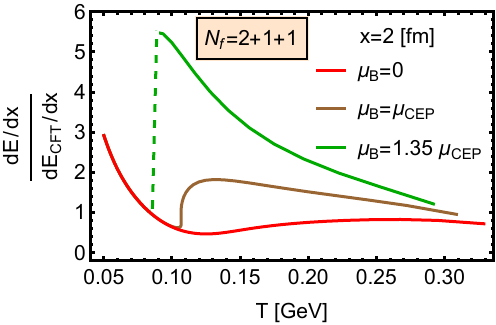}
\caption{Energy loss of light quark $\frac{dE/dx}{\left[dE/dx\right]_{CFT}}$ at finite temperature and baryon chemical potential with $x=2~\text{[fm]}$. The figures from top to bottom correspond to $N_f=2, 2+1, 2+1+1$ flavors, respectively. \textbf{Left panel}: the behavior of $\frac{dE/dx}{\left[dE/dx\right]_{CFT}}$ in the phase diagram. The CEP is denoted by green dots. \textbf{Right panel}: the temperature dependence of $\frac{dE/dx}{\left[dE/dx\right]_{CFT}}$ at fixed chemical potentials. The red, brown, and green lines correspond to $\mu_B=0,~ \mu_{CEP}~ \text{and}~ 1.35\mu_{CEP}$, respectively. The dashed lines in the figure indicate discontinuous jumps.}\label{fig4}
\end{figure}
%%%

In the previous section, we normalized the heavy quark energy loss by dividing by $T^2$, as $T^2$ and its conformal limit differ only by a velocity-related factor $-\frac{\pi}{2}\sqrt{\lambda_t} \frac{v}{\sqrt{1-v^2}}$, making the two normalizations equivalent. In contrast, the conformal limit of light-quark energy loss exhibits a more complex behavior regarding temperature and length scales. At small $x$, the conformal limit of the energy loss of the light quark $dE_{CFT}/dx$ is proportional to $T^2$, similar to the energy loss of elastic scattering of the pQCD \cite{Peshier:2006hi}. At intermediate $x$, it is proportional to $x T^3$, showing a path dependence similar to the radiative energy loss of pQCD \cite{Armesto:2011ht,Betz:2014cza}. At large $x$, it behaves as $x^2T^4$. To examine the asymptotic behavior of light-quark energy loss at high temperatures, we normalize it to the conformal limit in this section.

Fig.~\ref{fig4} illustrates the energy loss of light quarks at $x=2~\text{fm}$, which is also normalized by its conformal limit for $N_f=2,~2+1,~2+1+1$. Within the velocity range $v^2\leq 0.9$, the heavy and light quarks exhibit similar energy loss profiles across the phase diagram, as shown in the upper panels of Fig.~\ref{fig5} and Fig.~\ref{fig6}, as well as Fig.~\ref{fig4}. In the low-temperature and low-density phase, normalized energy loss $\frac{dE/dx}{dE_{CFT}/dx}$ decreases monotonically with increasing temperature while remaining largely insensitive to variations in chemical potential. Along the fixed chemical potential line and the phase boundary, the energy loss transitions from a decreasing trend to an increasing trend with temperature. It decreases again and forms a distinct peak structure upon crossing the phase boundary. Thus, the energy loss of light quarks near the phase boundary is enhanced, similar to that of heavy quarks. Normalized energy loss $\frac{dE/dx}{dE_{CFT}/dx}$ exhibits a non-monotonic behavior characterized by an initial decrease, followed by an increase and then a subsequent decrease. In the crossover region, this nonmonotonic variation occurs smoothly. Near the critical endpoint $\frac{dE/dx}{dE_{CFT}/dx}$ exhibits a rapid increase, while at the phase boundary of the first-order phase transition it undergoes a discontinuous upward jump. In the high-temperature phase, $\frac{dE/dx}{dE_{CFT}/dx}$ gradually decreases with increasing temperature and asymptotically approaches a constant value at sufficiently high temperatures, indicating that in the high-temperature limit the energy loss $d E/dx$ converges toward its conformal limit $d E_{CFT}/dx$. Fig.~\ref{fig4} shows that at a fixed temperature, $\frac{dE/dx}{dE_{CFT}/dx}$ exhibits monotonic growth with increasing chemical potential, with a particularly sharp increase observed near the phase boundary.

\begin{figure}[htbp]
\centering
\includegraphics[width=.4\textwidth]{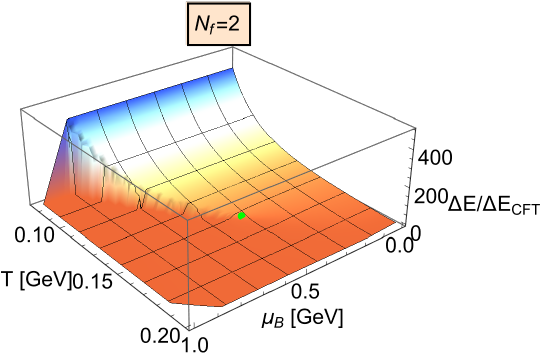}
\qquad
\includegraphics[width=.4\textwidth]{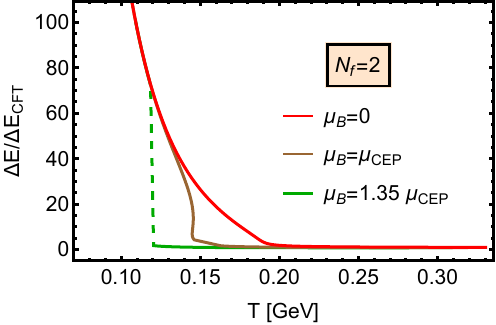}\\
\includegraphics[width=.4\textwidth]{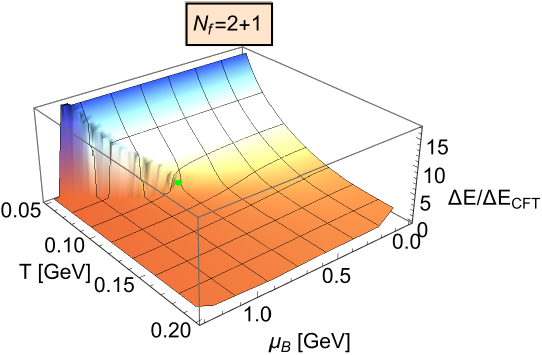}
\qquad
\includegraphics[width=.4\textwidth]{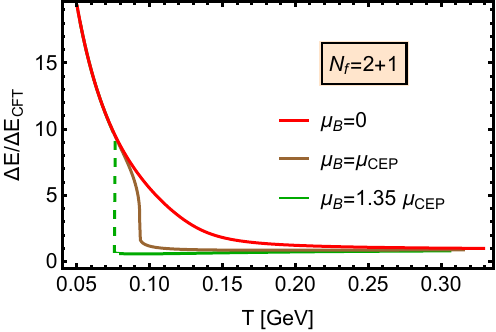}\\
\includegraphics[width=.4\textwidth]{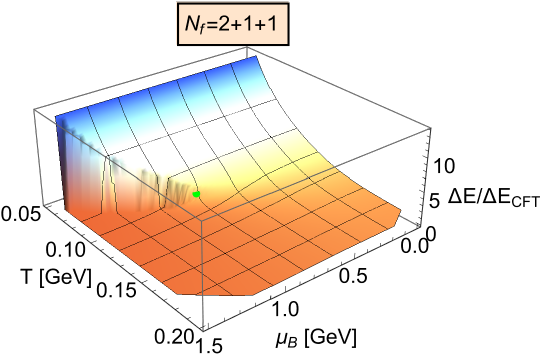}
\qquad
\includegraphics[width=.4\textwidth]{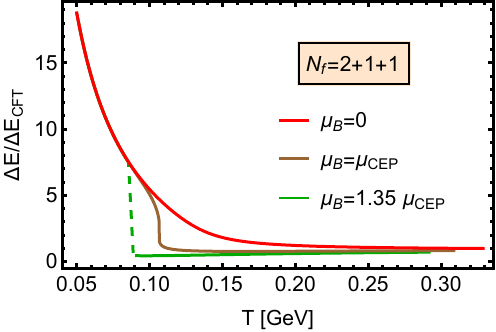}
\caption{Total energy loss of light quark $\Delta E/\Delta E_{CFT}$ at finite temperature and baryon chemical potential. The figures from top to bottom correspond to $N_f=2, 2+1, 2+1+1$ flavors, respectively. \textbf{Left panel}: the behavior of $\Delta E/\Delta E_{CFT}$ in the phase diagram. The CEP is denoted by green dots. \textbf{Right panel}: the temperature dependence of $\Delta E/\Delta E_{CFT}$ at fixed chemical potentials. The red, brown, and green lines correspond to $\mu_B=0,~ \mu_{CEP}~ \text{and}~ 1.35\mu_{CEP}$, respectively. The dashed lines in the figure indicate discontinuous jumps.}\label{fig3}
\end{figure}
%%%

Fig.~\ref{fig3} shows the normalized total energy loss of light quarks, expressed as a ratio to their conformal limit, for various quark flavors $N_f=2,~2+1,~2+1+1$. In the low-temperature and low-density phases, $\Delta E/\Delta E_{CFT}$ decreases with increasing temperature, but remains nearly invariant with respect to changes in chemical potential. In particular, at a fixed chemical potential, the normalized total energy loss $\Delta E/\Delta E_{CFT}$ decreases significantly with increasing temperature in the low-temperature phase. After crossing the phase boundary, the rate of this decrease gradually decreases. In contrast, the normalized total energy loss $\Delta E/\Delta E_{CFT}$ is reduced in the high-temperature phase. 

Compared to Fig.~\ref{fig3}, unlike the energy loss per unit length $dE/dx$, the total energy loss $\Delta E$ does not show a pronounced increase near the phase boundary. In the crossover region, the normalized total energy loss $\Delta E/\Delta E_{CFT}$ exhibits a smooth decrease with increasing temperature. Near the critical endpoint, a rapid decrease in $\Delta E/\Delta E_{CFT}$ is observed. At the phase boundary of the first-order phase transition, $\Delta E/\Delta E_{CFT}$ experiences a discontinuous jump and subsequent decrease.

\begin{figure}[htbp]
\centering
\includegraphics[width=.32\textwidth]{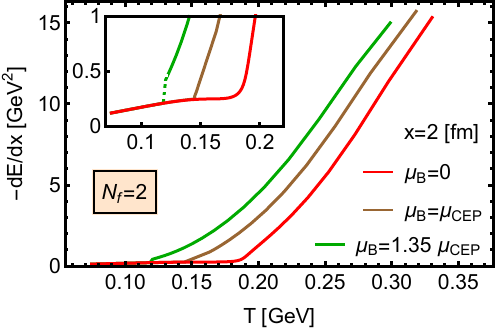}
\includegraphics[width=.32\textwidth]{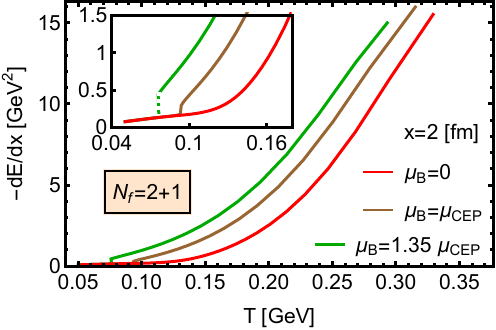}
\includegraphics[width=.32\textwidth]{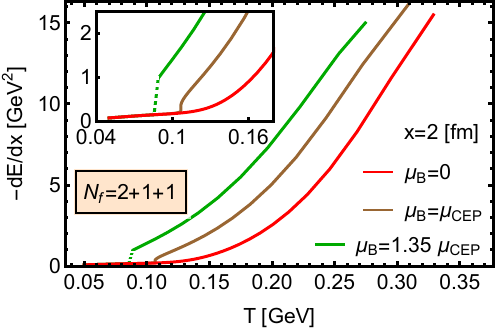}\\
\includegraphics[width=.32\textwidth]{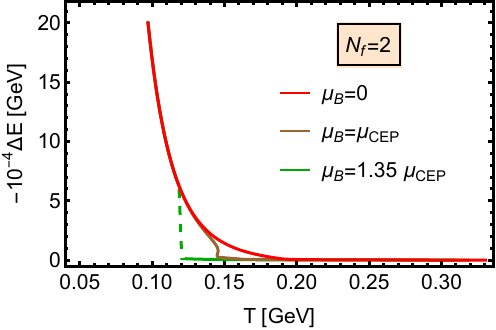}
\includegraphics[width=.32\textwidth]{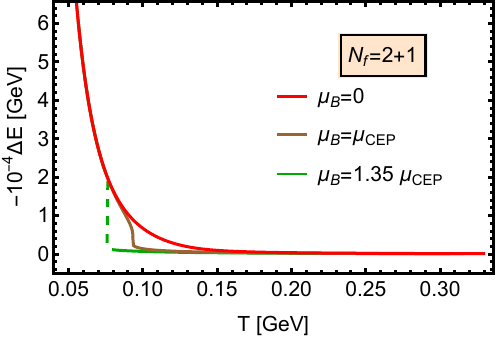}
\includegraphics[width=.32\textwidth]{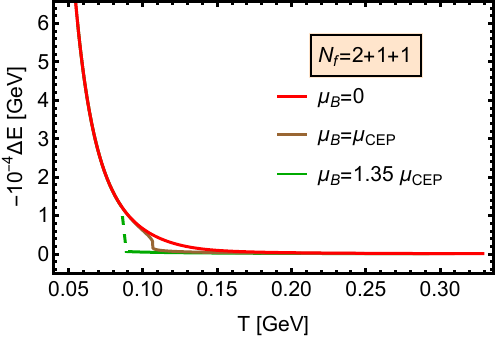}\\
\includegraphics[width=.32\textwidth]{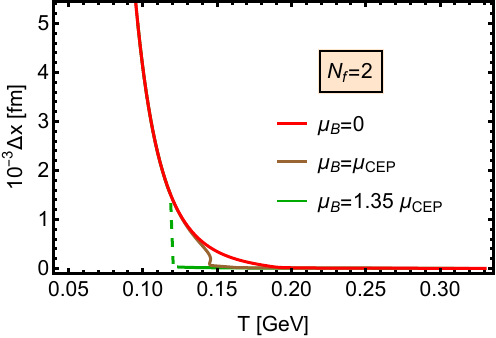}
\includegraphics[width=.32\textwidth]{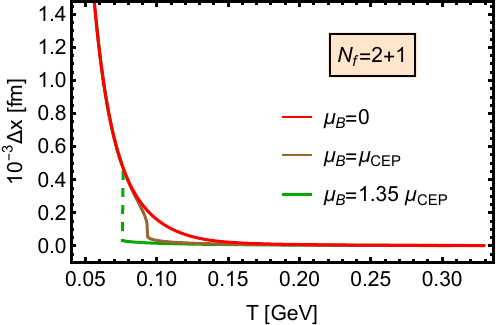}
\includegraphics[width=.32\textwidth]{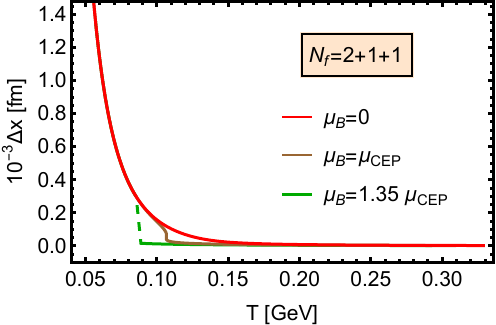}
\caption{Energy loss per unit length $-dE/dx$, total energy loss $\Delta E$ and the stopping distance $\Delta x$ at fixed baryon chemical potential. The figures from left to right correspond to $N_f=2, 2+1, 2+1+1$ flavors,  respectively. \textbf{Upper panel}: the temperature dependence of $-d E/dx$ at fixed baryon chemical potential.  \textbf{Middle panel}: the temperature dependence of $\Delta E$ at fixed baryon chemical potential. \textbf{Lower panel}: the temperature dependence of $\Delta x$ at fixed chemical potentials. The red, brown, and green lines correspond to $\mu_B=0,~ \mu_{CEP}~ \text{and}~ 1.35\mu_{CEP}$, respectively. The dashed lines in the figure indicate discontinuous jumps.}\label{figtt}
\end{figure}
%%%

To analyze why the energy loss per unit length $-dE/dx$ and the total energy loss $\Delta E$ behave differently, Fig.~\ref{figtt} shows how these measurements, along with the stopping distance $\Delta x$, change with temperature at constant chemical potentials.

As Fig.~\ref{figtt} indicates, $-dE/dx$ and $\Delta E$ respond oppositely to changes in temperature and chemical potential. The $-dE/dx$ increases with increasing temperature or chemical potential. In contrast, $\Delta E$ decreases with higher temperature and shows minimal change with chemical potential away from the phase boundary. However, near the phase boundary, the $\Delta E$ decreases as the chemical potential increases.

The stopping distance $\Delta x$ and the total energy loss $\Delta E$ behave similarly in response to changes in temperature and chemical potential. This suggests that the behavior of $\Delta E$ in the phase diagram depends mainly on the stopping distance $\Delta x$. Therefore, even though $-dE/dx$ increases with higher temperature or chemical potential, $\Delta E$ decreases due to the shorter stopping distance.

\section{Discussion and summary}

In this study, we investigate the jet quenching parameter and energy loss in holographic QCD models with $N_f=2,~2+1,~\text{and}~ 2+1+1$ flavors, which incorporate a critical end point (CEP) at finite baryon chemical potential. The temperature rescaled, dimensionless jet quenching parameter $\hat{q}/T^3$, light quark energy loss $\frac{dE/dx}{dE_{CFT}/dx}$, and heavy quark energy loss $F_{drag}/T^2$ are analyzed. These parameters, representing parton energy loss in the quark-gluon plasma (QGP) from different angles, display comparable trends on the phase diagram. Specifically, near the phase-transition temperature, these quantities exhibit a rapid increase. In the crossover region, this increase is smooth, while at the critical endpoint, the slope of the increase becomes infinite. In the first-order phase transition, the increase is discontinuous. As the temperature continues to increase, all three quantities approach their respective conformal limits and become independent of the chemical potential at very high temperatures. In the finite-temperature regime, these quantities increase with the chemical potential, indicating that partons experience greater resistance in a denser medium. We have also used holographic models with different flavors to compare the parton energy loss in QGP media with different flavors. It is found that, apart from slight differences in the phase boundary and the location of the CEP, the qualitative behavior of energy loss on the phase diagram is similar across different flavors, with only quantitative differences observed.

Our analysis of heavy-quark energy loss reveals consistent behavior in the phase diagram across a broad velocity range $v^2 \leq 0.9$, except in extreme relativistic cases $v \sim c$ where the behavior differs from low velocities. At constant temperature and chemical potential, heavy quark energy loss versus velocity aligns with Bjorken's results at high velocities and is approximately proportional to velocity at low velocities. Comparisons with the Bethe-Bloch formula and the Lindhard-Scharff-Schiott theory for electromagnetic interaction energy loss indicate that at low velocities, heavy quark energy loss is predominantly collisional, while at high velocities, it is mainly due to gluon radiation.

For light-quark energy loss, the behavior of the energy loss per unit length $dE/dx$ and the total energy loss $\Delta E$ in the phase diagram differs significantly. However, $\Delta E$ and the stopping distance $\Delta x$ exhibit a similar behavior. This implies that the stopping distance predominantly determines the total energy loss. Thus, even with an increase in energy loss per unit length at higher temperatures or chemical potentials, the total energy loss decreases because of the reduced stopping distance. It could be interesting to investigate the impacts of the rapid changes near CEP of those quantities on the experimental observables, but we will leave it to the future.

\begin{acknowledgments}
This work is supported by the National Natural Science Foundation of China (NSFC) under Grant Nos: 12235016, 12221005,12275108, the national Key Program for Science and Technology Research Development (2023YFB3002500) and the Natural Science Foundation of Henan Province of China (Grant No. 242300420235).
\end{acknowledgments}

%\bibliographystyle{utphys}
%\bibliography{ref.bib}

\begin{thebibliography}{10}
	
	\bibitem{Ashish2024}
	A.~P. for STAR~Collaboration, ``Precision measurement of net-proton number
	fluctuations in au+au collisions at rhic,''.
	
	\bibitem{STAR:2025zdq}
	{\bfseries STAR} Collaboration, ``{Precision Measurement of (Net-)proton Number
		Fluctuations in Au+Au Collisions at RHIC},''
	\href{http://arxiv.org/abs/2504.00817}{{\ttfamily arXiv:2504.00817
			[nucl-ex]}}.
	
	\bibitem{Shen:2024ptz}
	Y.~Shen, W.~Chen, X.-y. Wu, K.~Xu, and M.~Huang, ``{Net proton number cumulants
		from viscous hydrodynamics with an equation~of state including a critical end
		point},'' \href{http://dx.doi.org/10.1103/PhysRevC.111.014916}{{\em Phys.
			Rev. C} {\bfseries 111} no.~1, (2025) 014916},
	\href{http://arxiv.org/abs/2404.02397}{{\ttfamily arXiv:2404.02397
			[hep-ph]}}.
	
	\bibitem{Baier:1996sk}
	R.~Baier, Y.~L. Dokshitzer, A.~H. Mueller, S.~Peigne, and D.~Schiff,
	``{Radiative energy loss and p(T) broadening of high-energy partons in
		nuclei},'' \href{http://dx.doi.org/10.1016/S0550-3213(96)00581-0}{{\em Nucl.
			Phys. B} {\bfseries 484} (1997) 265--282},
	\href{http://arxiv.org/abs/hep-ph/9608322}{{\ttfamily arXiv:hep-ph/9608322}}.
	
	\bibitem{Shan-Liang:2023sdk}
	Z.~Shan-Liang, X.~Hong-Xi, and W.~En-Ke, ``{Jet quenching effect in
		relativistic heavy-ion collisions},''
	\href{http://dx.doi.org/10.7498/aps.72.20230993}{{\em Acta Phys. Sin.}
		{\bfseries 72} no.~20, (2023) 200304}.
	
	\bibitem{JET:2013cls}
	{\bfseries JET} Collaboration, K.~M. Burke {\em et~al.}, ``{Extracting the jet
		transport coefficient from jet quenching in high-energy heavy-ion
		collisions},'' \href{http://dx.doi.org/10.1103/PhysRevC.90.014909}{{\em Phys.
			Rev. C} {\bfseries 90} no.~1, (2014) 014909},
	\href{http://arxiv.org/abs/1312.5003}{{\ttfamily arXiv:1312.5003 [nucl-th]}}.
	
	\bibitem{Li:2014hja}
	D.~Li, J.~Liao, and M.~Huang, ``{Enhancement of jet quenching around phase
		transition: result from the dynamical holographic model},''
	\href{http://dx.doi.org/10.1103/PhysRevD.89.126006}{{\em Phys. Rev. D}
		{\bfseries 89} no.~12, (2014) 126006},
	\href{http://arxiv.org/abs/1401.2035}{{\ttfamily arXiv:1401.2035 [hep-ph]}}.
	
	\bibitem{Xu:2014tda}
	J.~Xu, J.~Liao, and M.~Gyulassy, ``{Consistency of Perfect Fluidity and Jet
		Quenching in semi-Quark-Gluon Monopole Plasmas},''
	\href{http://dx.doi.org/10.1088/0256-307X/32/9/092501}{{\em Chin. Phys.
			Lett.} {\bfseries 32} no.~9, (2015) 092501},
	\href{http://arxiv.org/abs/1411.3673}{{\ttfamily arXiv:1411.3673 [hep-ph]}}.
	
	\bibitem{Du:2022oaw}
	L.~Du and U.~Heinz, ``{Response of a baryon-charged medium to energetic
		partons},'' \href{http://dx.doi.org/10.1103/PhysRevC.106.034903}{{\em Phys.
			Rev. C} {\bfseries 106} no.~3, (2022) 034903},
	\href{http://arxiv.org/abs/2205.10262}{{\ttfamily arXiv:2205.10262
			[hep-ph]}}.
	
	\bibitem{Panero:2013pla}
	M.~Panero, K.~Rummukainen, and A.~Sch\"afer, ``{Lattice Study of the Jet
		Quenching Parameter},''
	\href{http://dx.doi.org/10.1103/PhysRevLett.112.162001}{{\em Phys. Rev.
			Lett.} {\bfseries 112} no.~16, (2014) 162001},
	\href{http://arxiv.org/abs/1307.5850}{{\ttfamily arXiv:1307.5850 [hep-ph]}}.
	
	\bibitem{JETSCAPE:2021ehl}
	{\bfseries JETSCAPE} Collaboration, S.~Cao {\em et~al.}, ``{Determining the jet
		transport coefficient $\hat{q}$ from inclusive hadron suppression measurements
		using Bayesian parameter estimation},''
	\href{http://dx.doi.org/10.1103/PhysRevC.104.024905}{{\em Phys. Rev. C}
		{\bfseries 104} no.~2, (2021) 024905},
	\href{http://arxiv.org/abs/2102.11337}{{\ttfamily arXiv:2102.11337
			[nucl-th]}}.
	
	\bibitem{Xie:2022ght}
	M.~Xie, W.~Ke, H.~Zhang, and X.-N. Wang, ``{Information-field-based global
		Bayesian inference of the jet transport coefficient},''
	\href{http://dx.doi.org/10.1103/PhysRevC.108.L011901}{{\em Phys. Rev. C}
		{\bfseries 108} no.~1, (2023) L011901},
	\href{http://arxiv.org/abs/2206.01340}{{\ttfamily arXiv:2206.01340
			[hep-ph]}}.
	
	\bibitem{Wu:2022vbu}
	J.~Wu, S.~Cao, and F.~Li, ``{Critical Opalescence and Its Impact on the Jet
		Quenching Parameter},''
	\href{http://dx.doi.org/10.1088/0256-307X/41/3/031202}{{\em Chin. Phys.
			Lett.} {\bfseries 41} no.~3, (2024) 031202},
	\href{http://arxiv.org/abs/2208.14297}{{\ttfamily arXiv:2208.14297
			[nucl-th]}}.
	
	\bibitem{Liu:2006ug}
	H.~Liu, K.~Rajagopal, and U.~A. Wiedemann, ``{Calculating the jet quenching
		parameter from AdS/CFT},''
	\href{http://dx.doi.org/10.1103/PhysRevLett.97.182301}{{\em Phys. Rev. Lett.}
		{\bfseries 97} (2006) 182301},
	\href{http://arxiv.org/abs/hep-ph/0605178}{{\ttfamily arXiv:hep-ph/0605178}}.
	
	\bibitem{Liu:2006he}
	H.~Liu, K.~Rajagopal, and U.~A. Wiedemann, ``{Wilson loops in heavy ion
		collisions and their calculation in AdS/CFT},''
	\href{http://dx.doi.org/10.1088/1126-6708/2007/03/066}{{\em JHEP} {\bfseries
			03} (2007) 066}, \href{http://arxiv.org/abs/hep-ph/0612168}{{\ttfamily
			arXiv:hep-ph/0612168}}.
	
	\bibitem{Li:2014dsa}
	D.~Li, S.~He, and M.~Huang, ``{Temperature dependent transport coefficients in
		a dynamical holographic QCD model},''
	\href{http://dx.doi.org/10.1007/JHEP06(2015)046}{{\em JHEP} {\bfseries 06}
		(2015) 046}, \href{http://arxiv.org/abs/1411.5332}{{\ttfamily arXiv:1411.5332
			[hep-ph]}}.
	
	\bibitem{Zhu:2020qyw}
	X.~Zhu and Z.-q. Zhang, ``{Jet quenching parameter from a soft wall AdS/QCD
		model},'' \href{http://dx.doi.org/10.1088/1674-1137/abab87}{{\em Chin. Phys.
			C} {\bfseries 44} no.~10, (2020) 105105},
	\href{http://arxiv.org/abs/2006.14324}{{\ttfamily arXiv:2006.14324
			[nucl-th]}}.
	
	\bibitem{Cao:2024jgt}
	X.~Cao and H.~Liu, ``{Impact of the phase transition on quark-gluon plasma with
		an extremely strong magnetic field in holographic QCD},''
	\href{http://dx.doi.org/10.1103/PhysRevD.111.026008}{{\em Phys. Rev. D}
		{\bfseries 111} no.~2, (2025) 026008},
	\href{http://arxiv.org/abs/2408.00467}{{\ttfamily arXiv:2408.00467
			[hep-th]}}.
	
	\bibitem{Chen:2022goa}
	Y.~Chen, D.~Li, and M.~Huang, ``{The dynamical holographic QCD method for
		hadron physics and QCD matter},''
	\href{http://dx.doi.org/10.1088/1572-9494/ac82ad}{{\em Commun. Theor. Phys.}
		{\bfseries 74} no.~9, (2022) 097201},
	\href{http://arxiv.org/abs/2206.00917}{{\ttfamily arXiv:2206.00917
			[hep-ph]}}.
	
	\bibitem{Rougemont:2020had}
	R.~Rougemont, ``{Jet quenching parameters in strongly coupled anisotropic
		plasmas in the presence of magnetic fields},''
	\href{http://dx.doi.org/10.1103/PhysRevD.102.034009}{{\em Phys. Rev. D}
		{\bfseries 102} no.~3, (2020) 034009},
	\href{http://arxiv.org/abs/2002.06725}{{\ttfamily arXiv:2002.06725
			[hep-ph]}}.
	
	\bibitem{Chen:2022obe}
	J.-X. Chen and D.-F. Hou, ``{Heavy quark potential and jet quenching parameter
		in a rotating D-instanton background},''
	\href{http://dx.doi.org/10.1140/epjc/s10052-024-12708-7}{{\em Eur. Phys. J.
			C} {\bfseries 84} no.~4, (2024) 447},
	\href{http://arxiv.org/abs/2202.00888}{{\ttfamily arXiv:2202.00888
			[hep-ph]}}.
	
	\bibitem{Ficnar:2014pmc}
	A.~Ficnar, \href{http://dx.doi.org/10.7916/D8M04417}{{\em {Holographic Jet
				Quenching}}}.
	\newblock PhD thesis, Columbia U., 2014.
	
	\bibitem{Jacobs:2004qv}
	P.~Jacobs and X.-N. Wang, ``{Matter in extremis: Ultrarelativistic nuclear
		collisions at RHIC},''
	\href{http://dx.doi.org/10.1016/j.ppnp.2004.09.001}{{\em Prog. Part. Nucl.
			Phys.} {\bfseries 54} (2005) 443--534},
	\href{http://arxiv.org/abs/hep-ph/0405125}{{\ttfamily arXiv:hep-ph/0405125}}.
	
	\bibitem{YousufJamal:2019pen}
	M.~Yousuf~Jamal and V.~Chandra, ``{Energy loss of heavy quarks in the isotropic
		collisional hot QCD medium},''
	\href{http://dx.doi.org/10.1140/epjc/s10052-019-7278-2}{{\em Eur. Phys. J. C}
		{\bfseries 79} no.~9, (2019) 761},
	\href{http://arxiv.org/abs/1907.12033}{{\ttfamily arXiv:1907.12033
			[nucl-th]}}.
	
	\bibitem{Gubser:2006bz}
	S.~S. Gubser, ``{Drag force in AdS/CFT},''
	\href{http://dx.doi.org/10.1103/PhysRevD.74.126005}{{\em Phys. Rev. D}
		{\bfseries 74} (2006) 126005},
	\href{http://arxiv.org/abs/hep-th/0605182}{{\ttfamily arXiv:hep-th/0605182}}.
	
	\bibitem{Herzog:2006gh}
	C.~P. Herzog, A.~Karch, P.~Kovtun, C.~Kozcaz, and L.~G. Yaffe, ``{Energy loss
		of a heavy quark moving through N=4 supersymmetric Yang-Mills plasma},''
	\href{http://dx.doi.org/10.1088/1126-6708/2006/07/013}{{\em JHEP} {\bfseries
			07} (2006) 013}, \href{http://arxiv.org/abs/hep-th/0605158}{{\ttfamily
			arXiv:hep-th/0605158}}.
	
	\bibitem{Gursoy:2009kk}
	U.~Gursoy, E.~Kiritsis, G.~Michalogiorgakis, and F.~Nitti, ``{Thermal Transport
		and Drag Force in Improved Holographic QCD},''
	\href{http://dx.doi.org/10.1088/1126-6708/2009/12/056}{{\em JHEP} {\bfseries
			12} (2009) 056}, \href{http://arxiv.org/abs/0906.1890}{{\ttfamily
			arXiv:0906.1890 [hep-ph]}}.
	
	\bibitem{Rougemont:2015wca}
	R.~Rougemont, A.~Ficnar, S.~Finazzo, and J.~Noronha, ``{Energy loss,
		equilibration, and thermodynamics of a baryon rich strongly coupled
		quark-gluon plasma},'' \href{http://dx.doi.org/10.1007/JHEP04(2016)102}{{\em
			JHEP} {\bfseries 04} (2016) 102},
	\href{http://arxiv.org/abs/1507.06556}{{\ttfamily arXiv:1507.06556
			[hep-th]}}.
	
	\bibitem{Zhu:2021nbl}
	Z.-R. Zhu, J.-X. Chen, X.-M. Liu, and D.~Hou, ``{Thermodynamics and energy loss
		in D dimensions from holographic QCD model},''
	\href{http://dx.doi.org/10.1140/epjc/s10052-022-10433-7}{{\em Eur. Phys. J.
			C} {\bfseries 82} no.~6, (2022) 560},
	\href{http://arxiv.org/abs/2109.02366}{{\ttfamily arXiv:2109.02366
			[hep-ph]}}.
	
	\bibitem{Grefa:2022sav}
	J.~Grefa, M.~Hippert, J.~Noronha, J.~Noronha-Hostler, I.~Portillo, C.~Ratti,
	and R.~Rougemont, ``{Transport coefficients of the quark-gluon plasma at the
		critical point and across the first-order line},''
	\href{http://dx.doi.org/10.1103/PhysRevD.106.034024}{{\em Phys. Rev. D}
		{\bfseries 106} no.~3, (2022) 034024},
	\href{http://arxiv.org/abs/2203.00139}{{\ttfamily arXiv:2203.00139
			[nucl-th]}}.
	
	\bibitem{Zhu:2023aaq}
	Z.-R. Zhu and D.~Hou, ``{Inverse magnetic catalysis and energy loss in a
		holographic QCD model},''
	\href{http://dx.doi.org/10.1103/PhysRevD.110.066010}{{\em Phys. Rev. D}
		{\bfseries 110} no.~6, (2024) 066010},
	\href{http://arxiv.org/abs/2305.12375}{{\ttfamily arXiv:2305.12375
			[hep-ph]}}.
	
	\bibitem{Chen:2024epd}
	B.~Chen, X.~Chen, X.~Li, Z.-R. Zhu, and K.~Zhou, ``{Exploring Transport
		Properties of Quark-Gluon Plasma with a Machine-Learning assisted Holographic
		Approach},'' \href{http://arxiv.org/abs/2404.18217}{{\ttfamily
			arXiv:2404.18217 [hep-ph]}}.
	
	\bibitem{Domurcukgul:2021qfe}
	T.~Domurcukgul and R.~Morad, ``{Holographic drag force in non-conformal
		plasma},'' \href{http://dx.doi.org/10.1140/epjc/s10052-022-10252-w}{{\em Eur.
			Phys. J. C} {\bfseries 82} no.~4, (2022) 304},
	\href{http://arxiv.org/abs/2108.10853}{{\ttfamily arXiv:2108.10853
			[hep-th]}}.
	
	\bibitem{Finazzo:2016mhm}
	S.~I. Finazzo, R.~Critelli, R.~Rougemont, and J.~Noronha, ``{Momentum transport
		in strongly coupled anisotropic plasmas in the presence of strong magnetic
		fields},'' \href{http://dx.doi.org/10.1103/PhysRevD.94.054020}{{\em Phys.
			Rev. D} {\bfseries 94} no.~5, (2016) 054020},
	\href{http://arxiv.org/abs/1605.06061}{{\ttfamily arXiv:1605.06061
			[hep-ph]}}. [Erratum: Phys.Rev.D 96, 019903 (2017)].
	
	\bibitem{Gubser:2008as}
	S.~S. Gubser, D.~R. Gulotta, S.~S. Pufu, and F.~D. Rocha, ``{Gluon energy loss
		in the gauge-string duality},''
	\href{http://dx.doi.org/10.1088/1126-6708/2008/10/052}{{\em JHEP} {\bfseries
			10} (2008) 052}, \href{http://arxiv.org/abs/0803.1470}{{\ttfamily
			arXiv:0803.1470 [hep-th]}}.
	
	\bibitem{Chesler:2008wd}
	P.~M. Chesler, K.~Jensen, and A.~Karch, ``{Jets in strongly-coupled N = 4 super
		Yang-Mills theory},''
	\href{http://dx.doi.org/10.1103/PhysRevD.79.025021}{{\em Phys. Rev. D}
		{\bfseries 79} (2009) 025021},
	\href{http://arxiv.org/abs/0804.3110}{{\ttfamily arXiv:0804.3110 [hep-th]}}.
	
	\bibitem{Ficnar:2012yu}
	A.~Ficnar, J.~Noronha, and M.~Gyulassy, ``{Falling Strings and Light Quark Jet
		Quenching at LHC},''
	\href{http://dx.doi.org/10.1016/j.nuclphysa.2012.12.030}{{\em Nucl. Phys. A}
		{\bfseries 910-911} (2013) 252--255},
	\href{http://arxiv.org/abs/1208.0305}{{\ttfamily arXiv:1208.0305 [hep-ph]}}.
	
	\bibitem{Ficnar:2013qxa}
	A.~Ficnar, S.~S. Gubser, and M.~Gyulassy, ``{Shooting String Holography of Jet
		Quenching at RHIC and LHC},''
	\href{http://dx.doi.org/10.1016/j.physletb.2014.10.016}{{\em Phys. Lett. B}
		{\bfseries 738} (2014) 464--471},
	\href{http://arxiv.org/abs/1311.6160}{{\ttfamily arXiv:1311.6160 [hep-ph]}}.
	
	\bibitem{Ficnar:2013wba}
	A.~Ficnar and S.~S. Gubser, ``{Finite momentum at string endpoints},''
	\href{http://dx.doi.org/10.1103/PhysRevD.89.026002}{{\em Phys. Rev. D}
		{\bfseries 89} no.~2, (2014) 026002},
	\href{http://arxiv.org/abs/1306.6648}{{\ttfamily arXiv:1306.6648 [hep-th]}}.
	
	\bibitem{Zhu:2019ujc}
	Z.-R. Zhu, S.-Q. Feng, Y.-F. Shi, and Y.~Zhong, ``{Energy loss of heavy and
		light quarks in holographic magnetized background},''
	\href{http://dx.doi.org/10.1103/PhysRevD.99.126001}{{\em Phys. Rev. D}
		{\bfseries 99} no.~12, (2019) 126001},
	\href{http://arxiv.org/abs/1901.09304}{{\ttfamily arXiv:1901.09304
			[hep-ph]}}.
	
	\bibitem{Chen:2024ckb}
	X.~Chen and M.~Huang, ``{Machine learning holographic black hole from lattice
		QCD equation of state},''
	\href{http://dx.doi.org/10.1103/PhysRevD.109.L051902}{{\em Phys. Rev. D}
		{\bfseries 109} no.~5, (2024) L051902},
	\href{http://arxiv.org/abs/2401.06417}{{\ttfamily arXiv:2401.06417
			[hep-ph]}}.
	
	\bibitem{Chen:2024mmd}
	X.~Chen and M.~Huang, ``{Flavor dependent Critical endpoint from holographic
		QCD through machine learning},''
	\href{http://arxiv.org/abs/2405.06179}{{\ttfamily arXiv:2405.06179
			[hep-ph]}}.
	
	\bibitem{Cai:2024eqa}
	R.-G. Cai, S.~He, L.~Li, and H.-A. Zeng, ``{Neural Ordinary Differential
		Equations for Mapping the Magnetic QCD Phase Diagram via Holography},''
	\href{http://arxiv.org/abs/2406.12772}{{\ttfamily arXiv:2406.12772
			[hep-th]}}.
	
	\bibitem{Fu:2024wkn}
	Q.~Fu, S.~He, L.~Li, and Z.~Li, ``{Revisiting holographic model for thermal and
		dense QCD with a critical point},''
	\href{http://arxiv.org/abs/2404.12109}{{\ttfamily arXiv:2404.12109
			[hep-ph]}}.
	
	\bibitem{Karch:2006pv}
	A.~Karch, E.~Katz, D.~T. Son, and M.~A. Stephanov, ``{Linear confinement and
		AdS/QCD},'' \href{http://dx.doi.org/10.1103/PhysRevD.74.015005}{{\em Phys.
			Rev. D} {\bfseries 74} (2006) 015005},
	\href{http://arxiv.org/abs/hep-ph/0602229}{{\ttfamily arXiv:hep-ph/0602229}}.
	
	\bibitem{Li:2011hp}
	D.~Li, S.~He, M.~Huang, and Q.-S. Yan, ``{Thermodynamics of deformed AdS$_5$
		model with a positive/negative quadratic correction in graviton-dilaton
		system},'' \href{http://dx.doi.org/10.1007/JHEP09(2011)041}{{\em JHEP}
		{\bfseries 09} (2011) 041}, \href{http://arxiv.org/abs/1103.5389}{{\ttfamily
			arXiv:1103.5389 [hep-th]}}.
	
	\bibitem{Li:2012ay}
	D.~Li, M.~Huang, and Q.-S. Yan, ``{A dynamical soft-wall holographic QCD model
		for chiral symmetry breaking and linear confinement},''
	\href{http://dx.doi.org/10.1140/epjc/s10052-013-2615-3}{{\em Eur. Phys. J. C}
		{\bfseries 73} (2013) 2615}, \href{http://arxiv.org/abs/1206.2824}{{\ttfamily
			arXiv:1206.2824 [hep-th]}}.
	
	\bibitem{Chelabi:2015gpc}
	K.~Chelabi, Z.~Fang, M.~Huang, D.~Li, and Y.-L. Wu, ``{Chiral Phase Transition
		in the Soft-Wall Model of AdS/QCD},''
	\href{http://dx.doi.org/10.1007/JHEP04(2016)036}{{\em JHEP} {\bfseries 04}
		(2016) 036}, \href{http://arxiv.org/abs/1512.06493}{{\ttfamily
			arXiv:1512.06493 [hep-ph]}}.
	
	\bibitem{Chelabi:2015cwn}
	K.~Chelabi, Z.~Fang, M.~Huang, D.~Li, and Y.-L. Wu, ``{Realization of chiral
		symmetry breaking and restoration in holographic QCD},''
	\href{http://dx.doi.org/10.1103/PhysRevD.93.101901}{{\em Phys. Rev. D}
		{\bfseries 93} no.~10, (2016) 101901},
	\href{http://arxiv.org/abs/1511.02721}{{\ttfamily arXiv:1511.02721
			[hep-ph]}}.
	
	\bibitem{Fang:2015ytf}
	Z.~Fang, S.~He, and D.~Li, ``{Chiral and Deconfining Phase Transitions from
		Holographic QCD Study},''
	\href{http://dx.doi.org/10.1016/j.nuclphysb.2016.04.003}{{\em Nucl. Phys. B}
		{\bfseries 907} (2016) 187--207},
	\href{http://arxiv.org/abs/1512.04062}{{\ttfamily arXiv:1512.04062
			[hep-ph]}}.
	
	\bibitem{Fang:2019lsz}
	Z.~Fang and Y.-L. Wu, ``{Equation of state and chiral transition in soft-wall
		AdS/QCD with more realistic gravitational background},''
	\href{http://arxiv.org/abs/1909.06917}{{\ttfamily arXiv:1909.06917
			[hep-ph]}}.
	
	\bibitem{DeWolfe:2010he}
	O.~DeWolfe, S.~S. Gubser, and C.~Rosen, ``{A holographic critical point},''
	\href{http://dx.doi.org/10.1103/PhysRevD.83.086005}{{\em Phys. Rev. D}
		{\bfseries 83} (2011) 086005},
	\href{http://arxiv.org/abs/1012.1864}{{\ttfamily arXiv:1012.1864 [hep-th]}}.
	
	\bibitem{DeWolfe:2011ts}
	O.~DeWolfe, S.~S. Gubser, and C.~Rosen, ``{Dynamic critical phenomena at a
		holographic critical point},''
	\href{http://dx.doi.org/10.1103/PhysRevD.84.126014}{{\em Phys. Rev. D}
		{\bfseries 84} (2011) 126014},
	\href{http://arxiv.org/abs/1108.2029}{{\ttfamily arXiv:1108.2029 [hep-th]}}.
	
	\bibitem{Cai:2012xh}
	R.-G. Cai, S.~He, and D.~Li, ``{A hQCD model and its phase diagram in
		Einstein-Maxwell-Dilaton system},''
	\href{http://dx.doi.org/10.1007/JHEP03(2012)033}{{\em JHEP} {\bfseries 03}
		(2012) 033}, \href{http://arxiv.org/abs/1201.0820}{{\ttfamily arXiv:1201.0820
			[hep-th]}}.
	
	\bibitem{Cai:2012eh}
	R.-G. Cai, S.~Chakrabortty, S.~He, and L.~Li, ``{Some aspects of QGP phase in a
		hQCD model},'' \href{http://dx.doi.org/10.1007/JHEP02(2013)068}{{\em JHEP}
		{\bfseries 02} (2013) 068}, \href{http://arxiv.org/abs/1209.4512}{{\ttfamily
			arXiv:1209.4512 [hep-th]}}.
	
	\bibitem{Finazzo:2013efa}
	S.~I. Finazzo and J.~Noronha, ``{Holographic calculation of the electric
		conductivity of the strongly coupled quark-gluon plasma near the
		deconfinement transition},''
	\href{http://dx.doi.org/10.1103/PhysRevD.89.106008}{{\em Phys. Rev. D}
		{\bfseries 89} no.~10, (2014) 106008},
	\href{http://arxiv.org/abs/1311.6675}{{\ttfamily arXiv:1311.6675 [hep-th]}}.
	
	\bibitem{Yang:2014bqa}
	Y.~Yang and P.-H. Yuan, ``{A Refined Holographic QCD Model and QCD Phase
		Structure},'' \href{http://dx.doi.org/10.1007/JHEP11(2014)149}{{\em JHEP}
		{\bfseries 11} (2014) 149}, \href{http://arxiv.org/abs/1406.1865}{{\ttfamily
			arXiv:1406.1865 [hep-th]}}.
	
	\bibitem{Critelli:2017oub}
	R.~Critelli, J.~Noronha, J.~Noronha-Hostler, I.~Portillo, C.~Ratti, and
	R.~Rougemont, ``{Critical point in the phase diagram of primordial
		quark-gluon matter from black hole physics},''
	\href{http://dx.doi.org/10.1103/PhysRevD.96.096026}{{\em Phys. Rev. D}
		{\bfseries 96} no.~9, (2017) 096026},
	\href{http://arxiv.org/abs/1706.00455}{{\ttfamily arXiv:1706.00455
			[nucl-th]}}.
	
	\bibitem{Li:2017ple}
	Z.~Li, Y.~Chen, D.~Li, and M.~Huang, ``{Locating the QCD critical end point
		through the peaked baryon number susceptibilities along the freeze-out
		line},'' \href{http://dx.doi.org/10.1088/1674-1137/42/1/013103}{{\em Chin.
			Phys. C} {\bfseries 42} no.~1, (2018) 013103},
	\href{http://arxiv.org/abs/1706.02238}{{\ttfamily arXiv:1706.02238
			[hep-ph]}}.
	
	\bibitem{Chen:2017cyc}
	Y.~Chen, M.~Huang, and Q.-S. Yan, ``{Gravitation waves from QCD and electroweak
		phase transitions},'' \href{http://dx.doi.org/10.1007/JHEP05(2018)178}{{\em
			JHEP} {\bfseries 05} (2018) 178},
	\href{http://arxiv.org/abs/1712.03470}{{\ttfamily arXiv:1712.03470
			[hep-ph]}}.
	
	\bibitem{Knaute:2017opk}
	J.~Knaute, R.~Yaresko, and B.~K\"ampfer, ``{Holographic QCD phase diagram with
		critical point from Einstein\textendash{}Maxwell-dilaton dynamics},''
	\href{http://dx.doi.org/10.1016/j.physletb.2018.01.053}{{\em Phys. Lett. B}
		{\bfseries 778} (2018) 419--425},
	\href{http://arxiv.org/abs/1702.06731}{{\ttfamily arXiv:1702.06731
			[hep-ph]}}.
	
	\bibitem{Fang:2018axm}
	Z.~Fang, Y.-L. Wu, and L.~Zhang, ``{Chiral phase transition and QCD phase
		diagram from AdS/QCD},''
	\href{http://dx.doi.org/10.1103/PhysRevD.99.034028}{{\em Phys. Rev. D}
		{\bfseries 99} no.~3, (2019) 034028},
	\href{http://arxiv.org/abs/1810.12525}{{\ttfamily arXiv:1810.12525
			[hep-ph]}}.
	
	\bibitem{Ballon-Bayona:2020xls}
	A.~Ballon-Bayona, H.~Boschi-Filho, E.~F. Capossoli, and D.~M. Rodrigues,
	``{Criticality from Einstein-Maxwell-dilaton holography at finite temperature
		and density},'' \href{http://dx.doi.org/10.1103/PhysRevD.102.126003}{{\em
			Phys. Rev. D} {\bfseries 102} no.~12, (2020) 126003},
	\href{http://arxiv.org/abs/2006.08810}{{\ttfamily arXiv:2006.08810
			[hep-th]}}.
	
	\bibitem{Li:2020hau}
	M.-W. Li, Y.~Yang, and P.-H. Yuan, ``{Analytic Study on Chiral Phase Transition
		in Holographic QCD},'' \href{http://dx.doi.org/10.1007/JHEP02(2021)055}{{\em
			JHEP} {\bfseries 02} (2021) 055},
	\href{http://arxiv.org/abs/2009.05694}{{\ttfamily arXiv:2009.05694
			[hep-th]}}.
	
	\bibitem{Grefa:2021qvt}
	J.~Grefa, J.~Noronha, J.~Noronha-Hostler, I.~Portillo, C.~Ratti, and
	R.~Rougemont, ``{Hot and dense quark-gluon plasma thermodynamics from
		holographic black holes},''
	\href{http://dx.doi.org/10.1103/PhysRevD.104.034002}{{\em Phys. Rev. D}
		{\bfseries 104} no.~3, (2021) 034002},
	\href{http://arxiv.org/abs/2102.12042}{{\ttfamily arXiv:2102.12042
			[nucl-th]}}.
	
	\bibitem{He:2022amv}
	S.~He, L.~Li, Z.~Li, and S.-J. Wang, ``{Gravitational Waves and Primordial
		Black Hole Productions from Gluodynamics},''
	\href{http://arxiv.org/abs/2210.14094}{{\ttfamily arXiv:2210.14094
			[hep-ph]}}.
	
	\bibitem{Grefa:2022fpu}
	J.~Grefa, M.~Hippert, J.~Noronha, J.~Noronha-Hostler, I.~Portillo, C.~Ratti,
	and R.~Rougemont, ``{QCD Equilibrium and Dynamical Properties from
		Holographic Black Holes},''
	\href{http://dx.doi.org/10.31349/SuplRevMexFis.3.040910}{{\em Rev. Mex. Fis.
			Suppl.} {\bfseries 3} no.~4, (2022) 040910},
	\href{http://arxiv.org/abs/2207.12564}{{\ttfamily arXiv:2207.12564
			[nucl-th]}}.
	
	\bibitem{Grefa:2023hmf}
	J.~Grefa, M.~Hippert, R.~Kunnawalkam~Elayavalli, J.~Noronha-Hostler,
	I.~Portillo, C.~Ratti, and R.~Rougemont, ``{Holographic transport
		coefficients and jet energy loss for the hot and dense quark-gluon plasma},''
	\href{http://dx.doi.org/10.1051/epjconf/202429614014}{{\em EPJ Web Conf.}
		{\bfseries 296} (2024) 14014},
	\href{http://arxiv.org/abs/2312.11449}{{\ttfamily arXiv:2312.11449
			[nucl-th]}}.
	
	\bibitem{He:2013qq}
	S.~He, S.-Y. Wu, Y.~Yang, and P.-H. Yuan, ``{Phase Structure in a Dynamical
		Soft-Wall Holographic QCD Model},''
	\href{http://dx.doi.org/10.1007/JHEP04(2013)093}{{\em JHEP} {\bfseries 04}
		(2013) 093}, \href{http://arxiv.org/abs/1301.0385}{{\ttfamily arXiv:1301.0385
			[hep-th]}}.
	
	\bibitem{Li:2017tdz}
	M.-W. Li, Y.~Yang, and P.-H. Yuan, ``{Approaching Confinement Structure for
		Light Quarks in a Holographic Soft Wall QCD Model},''
	\href{http://dx.doi.org/10.1103/PhysRevD.96.066013}{{\em Phys. Rev. D}
		{\bfseries 96} no.~6, (2017) 066013},
	\href{http://arxiv.org/abs/1703.09184}{{\ttfamily arXiv:1703.09184
			[hep-th]}}.
	
	\bibitem{Bjorken:1982tu}
	J.~D. Bjorken, ``{Energy Loss of Energetic Partons in Quark - Gluon Plasma:
		Possible Extinction of High p(t) Jets in Hadron - Hadron Collisions},''.
	
	\bibitem{SCHMIDT1987287}
	K.-H. Schmidt, E.~Hanelt, H.~Geissel, G.~Münzenberg, and J.~Dufour, ``The
	momentum-loss achromat — a new method for the isotopical separation of
	relativistic heavy ions,''
	\href{http://dx.doi.org/https://doi.org/10.1016/0168-9002(87)90092-1}{{\em
			Nuclear Instruments and Methods in Physics Research Section A: Accelerators,
			Spectrometers, Detectors and Associated Equipment} {\bfseries 260} no.~2,
		(1987) 287--303}.
	
	\bibitem{Lindhard:1961zz}
	J.~Lindhard and M.~Scharff, ``{Energy Dissipation by Ions in the kev Region},''
	\href{http://dx.doi.org/10.1103/PhysRev.124.128}{{\em Phys. Rev.} {\bfseries
			124} (1961) 128--130}.
	
	\bibitem{Bohr:1941zz}
	N.~Bohr, ``{Velocity-Range Relation for Fission Fragments},''
	\href{http://dx.doi.org/10.1103/PhysRev.59.270}{{\em Phys. Rev.} {\bfseries
			59} (1941) 270--275}.
	
	\bibitem{Stoecker:1994ud}
	H.~Stoecker {\em et~al.}, ``{Collective effects and nuclear stopping},''
	\href{http://dx.doi.org/10.1016/0375-9474(94)90605-X}{{\em Nucl. Phys. A}
		{\bfseries 566} (1994) 15C--26C}.
	
	\bibitem{NA49:1998gaz}
	{\bfseries NA49} Collaboration, H.~Appelshauser {\em et~al.}, ``{Baryon
		stopping and charged particle distributions in central Pb + Pb collisions at
		158-GeV per nucleon},''
	\href{http://dx.doi.org/10.1103/PhysRevLett.82.2471}{{\em Phys. Rev. Lett.}
		{\bfseries 82} (1999) 2471--2475},
	\href{http://arxiv.org/abs/nucl-ex/9810014}{{\ttfamily
			arXiv:nucl-ex/9810014}}.
	
	\bibitem{Peshier:2006hi}
	A.~Peshier, ``{The QCD collisional energy loss revised},''
	\href{http://dx.doi.org/10.1103/PhysRevLett.97.212301}{{\em Phys. Rev. Lett.}
		{\bfseries 97} (2006) 212301},
	\href{http://arxiv.org/abs/hep-ph/0605294}{{\ttfamily arXiv:hep-ph/0605294}}.
	
	\bibitem{Armesto:2011ht}
	N.~Armesto {\em et~al.}, ``{Comparison of Jet Quenching Formalisms for a
		Quark-Gluon Plasma 'Brick'},''
	\href{http://dx.doi.org/10.1103/PhysRevC.86.064904}{{\em Phys. Rev. C}
		{\bfseries 86} (2012) 064904},
	\href{http://arxiv.org/abs/1106.1106}{{\ttfamily arXiv:1106.1106 [hep-ph]}}.
	
	\bibitem{Betz:2014cza}
	B.~Betz and M.~Gyulassy, ``{Constraints on the Path-Length Dependence of Jet
		Quenching in Nuclear Collisions at RHIC and LHC},''
	\href{http://dx.doi.org/10.1007/JHEP10(2014)043}{{\em JHEP} {\bfseries 08}
		(2014) 090}, \href{http://arxiv.org/abs/1404.6378}{{\ttfamily arXiv:1404.6378
			[hep-ph]}}. [Erratum: JHEP 10, 043 (2014)].
	
\end{thebibliography}
\providecommand{\href}[2]{#2}\begingroup\raggedright\endgroup

\end{document}